# Metamorphic Testing of Deep Code Models: A Systematic Literature Review

ALI ASGARI, Delft University of Technology, The Netherlands
MILAN DE KONING, Delft University of Technology and JetBrains Research, The Netherlands
POURIA DERAKHSHANFAR, JetBrains Research, The Netherlands
ANNIBALE PANICHELLA, Delft University of Technology, The Netherlands

Large language models and deep learning models designed for code intelligence have revolutionized the software engineering field due to their ability to perform various code-related tasks. These models can process source code and software artifacts with high accuracy in tasks such as code completion, defect detection, and code summarization; therefore, they can potentially become an integral part of modern software engineering practices. Despite these capabilities, robustness remains a critical quality attribute for deep-code models as they may produce different results under varied and adversarial conditions (e.g., variable renaming). Metamorphic testing has become a widely used approach to evaluate models' robustness by applying semantic-preserving transformations to input programs and analyzing the stability of model outputs. While prior research has explored testing deep learning models, this systematic literature review focuses specifically on metamorphic testing for deep code models. By studying 45 primary papers, we analyze the transformations, techniques, and evaluation methods used to assess robustness. Our review summarizes the current landscape, identifying frequently evaluated models, programming tasks, datasets, target languages, and evaluation metrics, and highlights key challenges and future directions for advancing the field.

## 1 INTRODUCTION

In recent years, large language models for code (LLM4Code) have achieved remarkable performance across a range of software engineering tasks, reaching accuracy levels that make them increasingly viable for real-world adoption. These tasks include, but are not limited to, program repair [1], vulnerability detection [2, 3], code completion [4, 5], and clone detection [6, 7], among others. However, the practical applicability of LLM4Code depends not only on their performance on

Authors' addresses: Ali Asgari, a.asgari-2@tudelft.nl, Delft University of Technology, The Netherlands; Milan de Koning, milandekoning@tudelft.nl, Delft University of Technology and JetBrains Research, The Netherlands; Pouria Derakhshanfar, pouria.derakhshanfar@jetbrains.com, JetBrains Research, The Netherlands; Annibale Panichella, A.Panichella@tudelft.nl, Delft University of Technology, The Netherlands.

downstream tasks but also on the extent to which they meet non-functional requirements, such as robustness [8, 9, 10], sustainability [11, 12], security [13], and data privacy [14, 10]. For example, Yang et al. [15] have shown that LLM4Code often fails to satisfy many non-functional requirements, with robustness to adversarial code being particularly critical. Hammond [13] found that about 40% of codes generated by GitHub Copilot are affected by security vulnerabilities.

Among these requirements, *robustness* is especially critical since LLMs might not identify security vulnerabilities or repair bugs simply because developers have used different variable names or coding conventions [12]. Traditional testing methods, such as fuzzing [16] and test data generation [17], can generate new code snippets (input data) for LLMs but require model analysts to manually validate the correctness of the outputs. In particular, determining the expected output for a given input is time-consuming but also prone to human error, especially in the context of generated code, and often referred to as the *oracle problem* [18].

*Metamorphic testing* (MT) [19] offers a promising alternative by mitigating the oracle problem through the use of metamorphic relations—properties that describe how the output of a system (LLMs in our case) should behave when its input is systematically transformed. This technique is based on the premise that reasoning about relations between outputs of a model is often simpler than fully understanding the behavior of the model [19]. Metamorphic testing to deep code models is performed by applying metamorphic transformations to the model input (e.g., a code snippet) in a way that preserves the execution behavior of the code. The model under test is then assessed to produce the same prediction for both the original and transformed inputs. For example, a metamorphic transformation could involve exchanging a `for` loop with a `while` loop, which guarantees that the functionality of the code remains unchanged. By applying such transformations to a large collection of code snippets, the robustness of a model can be assessed by measuring how frequently the model's predictions change in response to transformed inputs.

Metamorphic testing has been applied in the literature to test programs from many application domains such as web services [20], image processing applications [21], and compiler testing [22]. The enduring relevance of metamorphic testing is further highlighted by numerous surveys and studies synthesizing and analyzing its applications across a wide range of domains [19, 23, 24, 25]. The most related survey is by Segura et al. [19], published in 2016, which discusses the usage of metamorphic testing in 12 domains, including classic machine learning (e.g., the WEKA library). However, this survey predates the advent of LLMs and does not consider the unique properties and challenges introduced by deep models of code. More recently, Rehman et al. [26] presented an overview of metamorphic testing applications to machine learning for various domains, such as protein function prediction [27] and six acoustic scene classifiers [28]. However, their focus is on the open challenges associated with metamorphic testing for supervised and unsupervised machine learning in a broader sense, rather than on LLMs for code-related tasks.

Due to the recent rise in research efforts and the increasing reported performance of LLMs, particularly in code-related tasks, researchers have proposed various recent techniques tailoring metamorphic testing specifically for LLMs applied to code (LLM4Code). Despite this growing body of research, there is a lack of a unified overview in the literature, and publications on metamorphic testing for LLM4Code are very fragmented. This is also due to the fact that authors used different terminologies when using metamorphic testing, e.g., counterfactual examples, adversarial attacks, or obfuscation techniques. At the time of writing this paper, there is no comprehensive survey or systematic literature review that synthesizes and analyzes these studies, making it difficult to assess the state of the art and identify new research directions. This paper addresses this gap by providing a systematic review of metamorphic testing for LLM4Code, offering a thorough analysis of existing work and highlighting future research opportunities.

Table 1. Main findings drawn from the survey for each Research question.

| Focus | Key Future Research Directions |
|---|---|
| **RQ1: Transformation Types** | (1) Add API-level rewrites, comment/whitespace edits, and full function-level refactors.<br>(2) Explore deeper, behaviour-preserving transformations (e.g., loop unrolling, memoisation).<br>(3) Quantify naturalness, readability, and compile-rate of transformed code.<br>(4) Develop AST-based, cross-language transformation libraries (e.g., for JS, Go, C#).<br>(5) Study compound transformations and their interactions.<br>(6) Release open corpora, reference implementations, and validators. |
| **RQ2: Application Techniques** | (1) Compare one-pass, gradient-based, evolutionary, sampling, and embedding-guided approaches using shared benchmarks.<br>(2) Create hybrid pipelines (e.g., seed genetic search with saliency gradients).<br>(3) Adapt MT to black-box LLMs using surrogate or transfer-based attacks.<br>(4) Use static and dynamic analysis to guide and constrain transformation selection.<br>(5) Report effectiveness and efficiency: queries, tokens changed, runtime.<br>(6) Open-source generators, fitness functions, and evaluation harnesses.<br>(7) Test cross-task and cross-language generalisation. |
| **RQ3: Downstream Tasks** | (1) Expand MT to new tasks: code repair, generation, completion, malware and memory issue detection.<br>(2) Emphasise safety-critical tasks (defect, vulnerability, malware detection).<br>(3) Evaluate MT across development pipelines (e.g., generation → compile → test).<br>(4) Build multilingual, multitask benchmark suites.<br>(5) Design task-specific metamorphic relations (e.g., API rewrites for translation). |
| **RQ4: Models under Test** | (1) Cover encoder-decoder and decoder-only models used in practice (e.g., CodeT5, StarCoder2, DeepSeek).<br>(2) Benchmark new checkpoints like Codestral, Magicoder, Phi-2.<br>(3) Create metamorphic relations for generation-oriented tasks.<br>(4) Compare robustness across multiple programming languages.<br>(5) Release patched models, transformation scripts, and robustness tables. |
| **RQ5: Programming Languages** | (1) Include JS/TS, C#, PHP, Go to reflect practical usage.<br>(2) Add functional (Haskell, OCaml) and system-level (Rust) languages.<br>(3) Design language-agnostic metamorphic relations.<br>(4) Release parsers and reusable transformation plugins per language.<br>(5) Report per-language robustness breakdowns in evaluations. |
| **RQ6: Datasets** | (1) Move beyond CodeSearchNet and BigCloneBench; use HumanEval, MBPP, StackEval, etc.<br>(2) Ensure coverage of underrepresented programming languages.<br>(3) Release hash lists and commit IDs to detect train–test leakage.<br>(4) Break down robustness by dataset and task.<br>(5) Share consistent source–transformed dataset splits. |
| **RQ7: Evaluation Metrics** | (1) Standardize metric names and cite exact formulas (e.g., F1, BLEU, ASR).<br>(2) Report uncertainty using confidence intervals or bootstrap bands.<br>(3) Pair task scores with semantic similarity and perturbation cost.<br>(4) Use significance tests (e.g., Wilcoxon, permutation) for comparisons.<br>(5) Share raw predictions to allow reanalysis and failure case diagnostics. |

In this study, we present a literature review that examines how metamorphic testing techniques are applied to evaluate the robustness of deep code models. A summary of our findings is provided in Table 1. We believe these results can play a significant role in guiding future research and development in this rapidly evolving area. This study offers the following principal contributions:

- We studied 45 papers as part of a systematic literature review (SLR) about metamorphic testing transformations of deep code models. All papers were carefully reviewed, analyzed, and classified following the methodology in Section 3.

- We classified different types of metamorphic transformations used in the literature, providing a taxonomy that organizes these transformations based on their characteristics, objectives, and applicability to various code-related tasks.
- We categorized various application techniques for MT in deep code models, emphasizing that not only the transformation type, but also the *application method* —such as one-pass, sampling-based, gradient-guided, or evolutionary— is critical to robustness assessment.
- We identified and summarized the deep code models, programming tasks, datasets, evaluation metrics, and target programming languages that are most frequently evaluated through metamorphic testing. This analysis provides insights into current trends in the application of metamorphic testing to code-related tasks.
- We outlined the limitations faced in metamorphic testing for deep code models and proposed future directions for researchers, paying close attention to current gaps in the field.

Section 2 summarizes the related work and introduces the core principles of deep code models, metamorphic transformations, and their applications in more detail. Section 3 describes the research questions, the methodology we followed to identify the relevant research papers, the inclusion and exclusion criteria for our study, and the data collected from each primary study. Our findings are described and summarized in Section 4 while Section Section 5 discusses the open challenges and future research directions we have identified. Finally, our conclusions are presented in Section 7. All data and materials related to this project are publicly available at:

github.com/SERG-Delft/Metamorphic-Testing-of-Deep-Code-Models

## 2  BACKGROUND & RELATED WORKS

### 2.1  Deep Code Models

In recent years, researchers in the software engineering community have used deep code models to (semi)automate various code-related tasks such as program repair [1], vulnerability detection [2, 3], and code completion [4, 5]. Deep code models can be categorized into three broad categories: (1) Natural language models for code, (2) Pre-trained code models, and (3) General-purpose large language models (LLMs).

*Natural Language Models for Code* focus on learning compact representations of code snippets and are directly trained on code data. These models extract embeddings from code snippets by incorporating structural information such as Abstract Syntax Trees (ASTs) and control flow graphs. Representative models in this category code2vec [29] and code2seq [30]. code2vec [29] uses a neural network to represent code snippets as fixed-length vectors, while graph-based models like GGNN (Gated Graph Neural Networks) [31] model code as graphs. Models in this category are trained and tuned directly for specific software engineering tasks.

*Pretrained Code Models* are large language models (LLMs) pre-trained on large corpora of software datasets, including GitHub and Stack Overflow, using self-supervised learning. For instance, CodeBERT [32] is a bimodal transformer model trained on both code and natural language descriptions, while CodeT5 [33] extends the T5 architecture for code. These models are pre-trained using masked language modeling or causal language modeling. Masked language modeling (MLM) consists of masking some tokens in the input sequence and optimizing the model to predict the masked tokens based on surrounding (unmasked) tokens. Instead, causal language modeling (CLM) is an autoregressive method that optimizes the model to predict the next token in a sequence given the previous tokens. After pre-training, these models are fine-tuned on task-specific datasets to adapt to specialized downstream tasks. For example, DeepTyper [34] is fine-tuned for type inference

*General-Purpose Large Language Models (LLMs)* are trained on multiple types of corpora beyond source code, including scientific papers, books, and broader natural language documents. These models, which often contain billions of parameters, can perform a wide range of natural language processing tasks without being fine-tuned for specific applications. These models can, therefore, perform downstream tasks other than those related to code but can also be further specialized to specific tasks via prompt engineering. Examples of general-purpose LLMs are GPT-4 [35], ChatGP [36], and LLaMA [37].

## 2.2 Metamorphic Testing

Metamorphic testing was first introduced by Chen et al. [38] as a solution to the *oracle problem*, where determining the correctness of outputs for given inputs is infeasible [18]. Instead of directly verifying correctness, metamorphic testing reasons about relationships between inputs and outputs [19]. For example, to verify a system computing the sine function, one can test the relation:

$$x_1 = \pi - x_2 \rightarrow \sin(x_1) = \sin(x_2).$$

This approach identifies errors if the system violates expected relationships, even without access to a test oracle. Applying metamorphic testing to deep code models involves semantic-preserving transformations of code snippets. These transformations, such as replacing a for loop with an equivalent while loop, maintain the original program's execution behavior. Robustness is then evaluated by measuring whether the model's predictions remain consistent between the original and transformed samples [9]. While metamorphic testing is widely applied in other domains such as web services, simulation and modeling, computer graphics, embedded systems, and machine learning [19], it presents unique challenges for code models due to strict syntactic and grammatical constraints. For instance, transformations must preserve both syntax and semantics, making the automatic generation and application of these transformations complex. Additionally, transformations may alter input-output behavior in task-specific ways. Bielik and Vechev [39] demonstrated transformations for type inference tasks, such as changing integer literals, where the altered output remains valid for the task. This highlights the adaptability of metamorphic testing to different contexts. Currently, metamorphic testing is primarily used to evaluate robustness, as reflected in most primary studies analyzed in this review. Its focus on semantic-preserving transformations provides a structured approach to robustness evaluation for deep code models, addressing critical challenges in software engineering and AI for code.

Figure 1 illustrates the core principle of metamorphic testing applied to deep-code models. Starting from a source code input, a *Transformation Selector/ Test Case Generator* first determines which transformations to apply. This component may use one-pass, sampling-based, or gradient-informed selection strategies. Guided by this, a suite of semantic-preserving transformations (e.g., variable renaming, dead code insertion, or statement reordering) is then applied to produce logically equivalent code variants. These variants, along with the original code, are passed to the System Under Test (SUT), typically a deep code model such as CodeBERT or CodeT5. The outputs are then compared: if the model yields inconsistent predictions across semantically equivalent inputs, the test is considered failed. This approach enables robustness evaluation without requiring oracle labels, relying instead on the semantic equivalence of transformed inputs as an implicit oracle.

## 2.3 Scope of Quality Attributes in Metamorphic Testing

While metamorphic testing (MT) has been widely applied to assess robustness [40, 41], its conceptual framework is not limited to this single quality attribute. In principle, MT can be extended to evaluate a broader set of non-functional properties, including fairness [42, 43, 44, 45, 46], explainability, and hallucination. However, in current practice, **especially within deep code model evaluation**, the use of MT remains predominantly centered on models' robustness. This trend is reflected in our review: all 45 primary studies analyzed employ MT to test models' resilience to input perturbations or semantic-preserving transformations, rather than to assess fairness or other attributes. The broader potential of MT to support quality evaluation beyond robustness remains underexplored, and we encourage future work to investigate how MT techniques can be extended or adapted to address fairness, hallucination detection, and explainability in LLM-based systems.

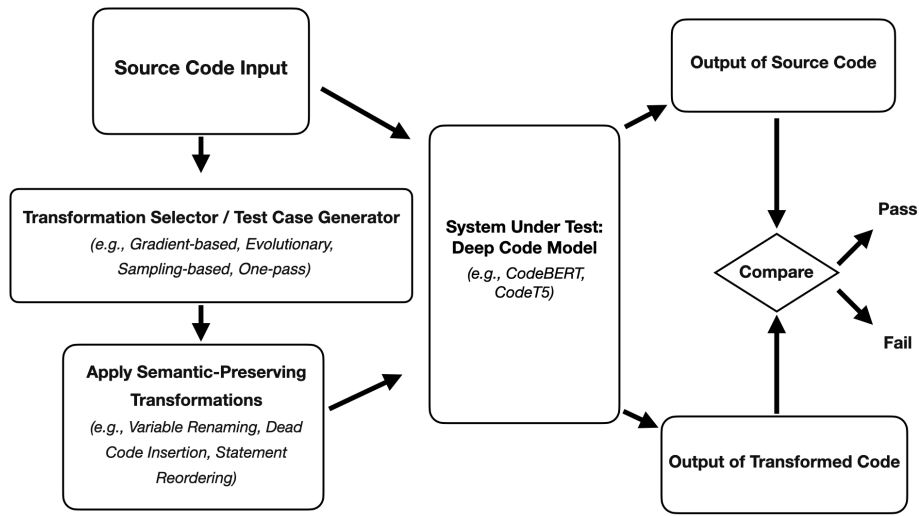

Fig. 1. Metamorphic Testing for Deep Code Models

### 2.4 Related Work

Researchers have investigated the robustness of deep code models in terms of their ability to maintain constant and correct performance in relation to minor perturbations in inputs [47]. Robustness failure can cause significant issues, such as generating inconsistent output for users with varying coding styles or English proficiency, compromising reliability [48]. In high-stakes cases, such as in CI/CD pipelines responsible for detecting compromised vulnerabilities, non-robust models can miss changed vulnerabilities, allowing potentially dangerous code to pass undetected. Robustness testing is generally divided between white-box and black-box methodologies [47, 49]; white-box methodologies require full access to model structure, parameters, and training data, whereas black-box methodologies work in the absence of such access [49].

Despite this large research effort in assessing the robustness of deep code models and the application of metamorphic testing, yet significant gaps remain. Li et al. [50] carried out a survey of the robustness of deep code models focusing on adversarial training and repair techniques for improving robustness. While their survey provides an extensive overview of adversarial methods, it does not look at the transformations and methods used to test robustness in different studies. Similarly, Qu et al. [51] studied the evaluation of robustness through semantic preservation transformations or adversarial attacks, but their survey, however largely investing in the topic, did not provide complete analyses or categorizations of the transformations. For instance, identifier renaming, a popular transformation, is repeatedly defined with inconsistent levels of elaboration across various studies, as illustrated in Table 5. Furthermore, Qu et al. omitted many relevant studies included in this work, leaving an enormous gap in coverage. Rabin et al. [52] extensively explored deep code models' properties and quality attributes, including robustness. However, the main focus was on methodologies for understanding these models through experimental evaluations and implementations rather than carrying out a survey or systematic literature review. Yang et al. [47] and Hou et al. [53] provide broader systematic literature reviews focusing on many aspects and different dimensions of Large Language Models for Code. Robustness, security, privacy, and explainability are explored in their study. However, their scope extends beyond robustness, and they do not focus only on testing methods.

Earlier surveys such as [23] offer a broad overview of metamorphic testing by discussing its main ideas, general challenges, and future possibilities. However, our study addresses a more specific and timely topic: the use of metamorphic testing for deep code models. Unlike previous reviews, we systematically examine which transformations are used, how they are applied, and which models, tasks, datasets, languages, and evaluation metrics are involved. This level of technical and empirical detail has not been covered before. In addition, our discussion of challenges and future directions is based directly on the patterns and gaps we observed in the literature, rather than on general assumptions. This makes our review the first detailed and data-driven overview of how metamorphic testing is currently applied to the evaluation of deep code models.

*In contrast to these prior works, our study narrows its focus specifically to metamorphic testing and provides an in-depth analysis of its application in robustness evaluation for deep code models. We systematically categorize transformation types, clarify inconsistent terminology (e.g., "metamorphic testing" vs. "counterfactual examples"), and incorporate a broader and more up-to-date set of studies. This enables us to fill key gaps in the literature and offer the first comprehensive review of metamorphic testing practices for LLM4Code.*

## 3 METHODOLOGY

In conducting our systematic review, we have followed a rigorous methodology guided by frameworks developed by Webster et al. [54] and Kitchenham et al. [55]. Webster et al. and Kitchenham et al.'s approaches complement one another; Webster et al. promote the integration of strands of studies in a range of subjects, and Kitchenham et al. present a specific technique for conducting systematic reviews specifically in software engineering. Our methodology is based on a comprehensive and unbiased analysis of the literature on metamorphic testing for deep code models.

### 3.1 Research questions

To gain a better overview of the field of metamorphic testing of deep code models, we formulated the following research questions.

- **RQ1:** *What types of metamorphic transformations have been used to assess the robustness of deep code models?*
- **RQ2:** *What techniques have been used to apply transformations in robustness evaluation?*
- **RQ3:** *Which programming-related tasks have been evaluated using metamorphic testing?*
- **RQ4:** *What deep code models have been tested through metamorphic testing?*
- **RQ5:** *What programming languages appear in input code used for metamorphic testing of deep code models?*
- **RQ6:** *What datasets have been used to generate metamorphic transformations and assess model performance?*
- **RQ7:** *What evaluation metrics are used to assess the effectiveness of metamorphic transformations on deep code models?*

Figure 2 presents the conceptual framework that underpins our systematic literature review on metamorphic testing for deep code models. The framework is structured into seven analytical dimensions corresponding to our research questions (RQ1–RQ7). We begin with an analysis of the types of metamorphic transformations (e.g., renaming variables, inserting dead code) in RQ1, followed by an exploration of transformation application techniques (e.g., one-pass strategies or search-based methods) in RQ2. RQs 3 through 6 examine the broader evaluation context, including the targeted tasks (e.g., summarization, defect detection), deep code models (e.g., CodeBERT, CodeT5), programming languages, and datasets. Finally, RQ7 investigates the evaluation metrics (e.g., Attack Success Rate, BLEU, Accuracy) used to assess model robustness. This structured framework provides a holistic analysis of the metamorphic testing pipeline and its components, helping to identify prevailing practices and highlight opportunities for future improvements.

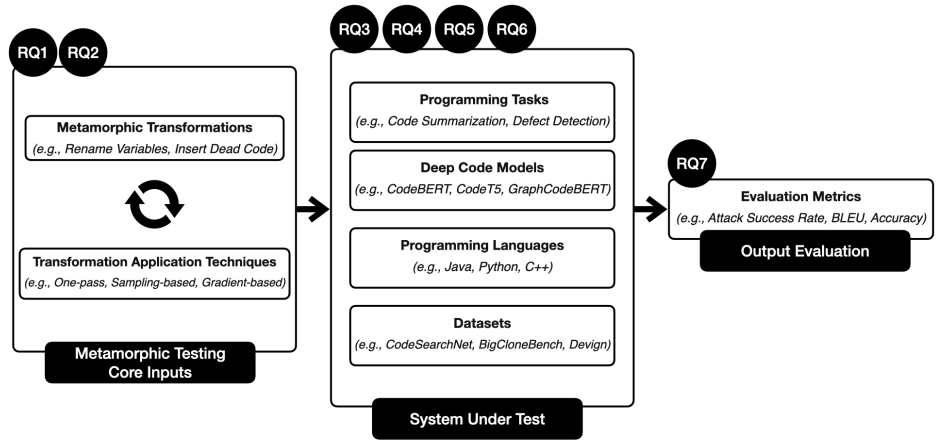

Fig. 2. Metamorphic Testing for Deep Code Models Framework

### 3.2 Inclusion and Exclusion Criteria

For this literature review, we focus on papers that meet specific inclusion and exclusion criteria to ensure the relevance and quality of our study. The primary focus is on research applying metamorphic or label-preserving transformations that do not alter a code snippet's behavior with respect to the task. These transformations must then be used to evaluate the robustness of deep code models.

For clarity, the inclusion and exclusion criteria are summarized in Table 2.

### 3.3 Publishing Venues

As illustrated in Figure 3, the number of publications on metamorphic testing for deep code models has increased significantly since 2020, with a peak in 2023. Most of the papers (58.5%) appeared in conference proceedings, followed by arXiv preprints (22.0%) and journal articles (19.5%), highlighting the field's rapid development and preference for fast dissemination. Figure 4 presents the distribution of individual venues. The most common venues include arXiv (8 papers), ICSE (International Conference on Software Engineering), SANER (International Conference on Software Analysis, Evolution and Reengineering), and ICML (International Conference on Machine Learning). Other notable conferences include ASE (Automated Software Engineering), GECCO (Genetic and Evolutionary Computation Conference), and ACL (Annual Meeting of the Association for Computational Linguistics). For journals, TOSEM (ACM Transactions on Software Engineering and Methodology), TIFS (IEEE Transactions on Information Forensics and Security), and IST (Information and Software Technology) are among the most represented. This distribution confirms that research on metamorphic testing in deep code models spans across software engineering, machine learning, and security communities.

Table 2. Inclusion and exclusion criteria for the systematic literature review (2019 to September 10th, 2024).

| Category | Description |
|---|---|
| **Inclusion Criteria** | • Papers applying metamorphic or label-preserving transformations that do not alter the semantics of code snippets. |
| | • Papers where modified code snippets are used for robustness evaluations of deep code models. |
| | • Research focusing on robustness testing for code-related tasks, such as summarization, completion, and defect detection, where adversarial code modifications or metamorphic testing transformations are applied. |
| **Exclusion Criteria** | • Papers focusing solely on data augmentation for deep code models. |
| | • Research using semantic-altering (i.e., non-metamorphic) transformations as adversarial attacks. |
| | • Studies applying metamorphic transformations to systems other than deep code models. |
| | • Publications written in languages other than English. |

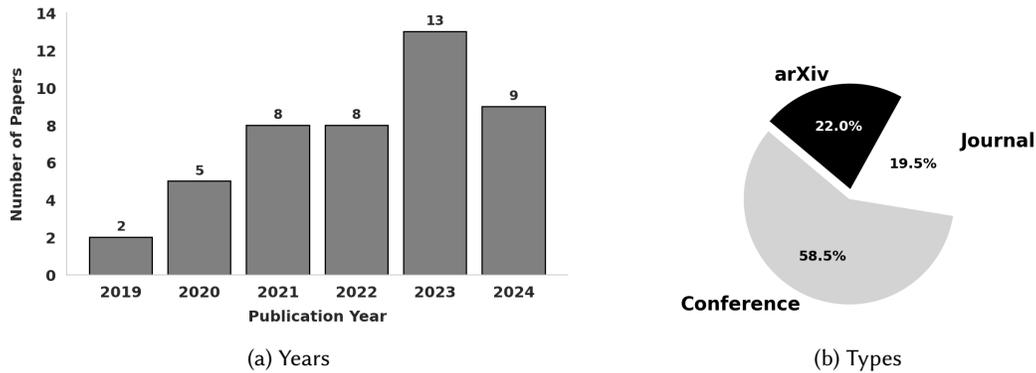

Fig. 3. Overview of publication years and publication types.

Table 3. Query Terms

| Concept | Related Terms/Keywords |
|---|---|
| Metamorphic testing | "Metamorphic", "Counterfactual", "Obfuscat*", "Fuzz*", "Adversar*", "Perturb*" |
| Code models | "Code models", "Software Engineering ML", "Software Engineering AI", "ML4SE", "AI4SE", "Program analysis tool*", "LLM4Code" |

## 3.4 Sources and Search strategy

To identify papers relevant to metamorphic testing and deep code models, we constructed a search query that ensured both concepts were present in the title, keywords, or abstract. The query was based on the concepts and identified synonyms/keywords shown in Table 3.

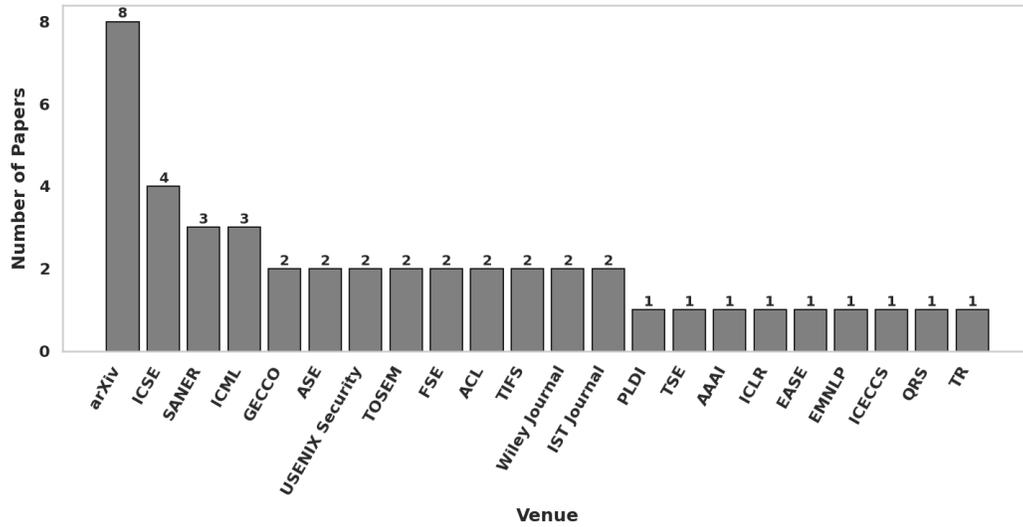

Fig. 4. Publishing Venues

Table 4. Databases searched, number of retrieved and selected primary studies

| Database | Retrieved Papers | Selected Primary Studies |
|---|---|---|
| ACM Digital Library | 326 | 11 |
| IEEE Xplore | 36 | 11 |
| Wiley Online Library | 39 | 2 |
| Elsevier ScienceDirect | 54 | 1 |
| Google Scholar[†] | – | 20 |
| **Total** | **455** | **45** |

[†]The query produces thousands of hits; the table lists only the 20 papers that fulfilled the inclusion criteria.

We constructed a search query by adding a disjunction(OR) between each of the related terms and a conjunction(AND) between the concepts as follows:

```
("Metamorphic" OR "Counterfactual" OR ...) AND
("Code models" OR "Software Engineering ML" OR ...)
```

This query was applied across multiple academic databases, and the number of primary studies retrieved from each source is summarized in Table 4. The search covered papers published from 2019 to September 10th, 2024.

We began our study by applying a structured search query across four major academic databases: ACM Digital Library, IEEE Xplore, Wiley Online Library, and Elsevier ScienceDirect. The initial query retrieved a total of 455 candidate records (Table 4, Figure 5): 326 from ACM, 36 from IEEE, 39 from Wiley, and 54 from Elsevier. Each retrieved paper underwent a manual quality assessment based on the criteria in Table 2. Papers clearly unrelated to metamorphic testing or deep code models were excluded. This filtering process resulted in 25 relevant primary studies. To broaden our coverage, we performed a targeted follow-up search on Google Scholar using a similar query, which had a large result set. From this set, we manually selected 20 additional papers that met our inclusion criteria, leading to a final total of 45 primary studies. While our search strategy was comprehensive, we acknowledge that some relevant studies may still have been missed due to naming variations and limited indexing in certain venues. Nevertheless, we are confident that our review covers the most significant and representative contributions in this domain.

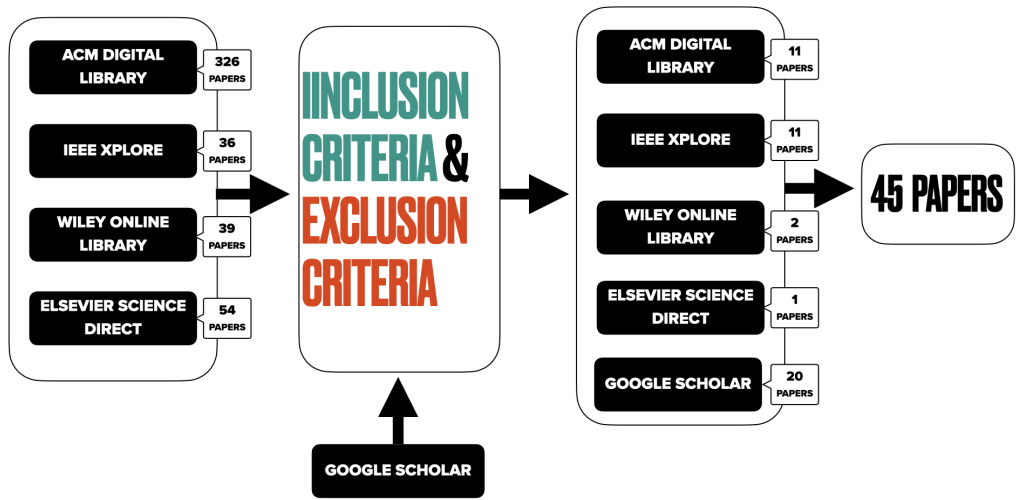

Fig. 5. Methodology Overflow

## 3.5 Data collection

To answer the research questions, we systematically recorded key information from each paper:

- *Types of transformations used*: We manually analyzed and categorized the metamorphic transformations applied to input code snippets, such as variable renaming, loop restructuring, and control flow modifications. This analysis helps identify common patterns in robustness evaluations.
- *Strategy for applying transformations*: We documented how metamorphic transformations were applied, whether randomly, systematically, or based on specific heuristics. Understanding these techniques provides insight into how test cases are generated and robustness is assessed.
- *Downstream tasks*: We recorded the programming tasks targeted by metamorphic testing, such as code summarization, completion, defect detection, and translation. This helps identify the most frequently evaluated tasks and potential gaps in the literature.
- *Models under test*: We documented the deep code models evaluated in each study, including their architectures (e.g., transformer-based models like CodeBERT or CodeT5) and training methodologies. This analysis reveals which models are most commonly tested and how their robustness is assessed.
- *Downstream languages*: We recorded the programming languages used in test datasets, such as Python, Java, and C++, to identify biases or gaps in language coverage in metamorphic robustness evaluations.

- *Test datasets*: We documented the datasets used for evaluation, including their size, source, and characteristics, to provide context for experimental setups and assess the generalizability of results.
- *Evaluation metrics*: We recorded the metrics used to assess model performance under metamorphic transformations. This includes both traditional metrics (e.g., Accuracy, BLEU) and robustness specific indicators. Documenting these metrics helps us understand the criteria used in prior studies and enables a comparison of evaluation standards across different experimental settings.

The first two authors classified each study based on predefined research questions and associated metrics, identifying 121 unique transformations across all papers, categorized into 26 datasets, 44 models, and 15 task transformations. They came into 17 disagreements or doubts. In cases of disagreement, discussions were held among all authors to resolve inconsistencies and reach a consensus, to alleviate potential biases and inconsistencies in the manual classification and data validation process.

## 4 RESULTS

This section summarizes our findings regarding types of metamorphic transformations, techniques, target models, and downstream tasks.

### 4.1 RQ1: Types of Metamorphic Transformations

This section describes the transformations used by each primary study identified using the methodology described in the previous sections, thus addressing our RQ1. In total, we found 121 transformations in the primary studies, and we provide the full list of transformations along with descriptions and examples in the supplementary material. We categorize the transformations based on the categories defined by Liu et al. [56] and Quiring et al. [57]. Some categories were added to accommodate groups of transformations that did not fit into any categories defined by the aforementioned works. Table 5 shows the number of transformation types in each category. The categories are as follows:

*4.1.1 Trivial Transformations.* Trivial code transformations are simple modifications that alter low-level aspects of a program without affecting its basic functionality. These transformations are common in program analysis and testing, as they were observed in 36% of the studies reviewed in our literature review. Below, we categorize and describe some of the most frequently encountered transformations, along with examples from prior research.

*Permute Statement.* The Permute Statement transformation involves rearranging statements in a program while preserving their original intent and purpose. This transformation captures the simplicity of straightforward, trivial modifications. For instance, Springer et al. [58] demonstrate how rearranging statements maintains program semantics. Similarly, Rabin et al. [59] use a syntactic permutation of statements to ensure semantic preservation. In a follow-up study, Rabin et al. [60] further explore this by swapping independent statements and adding minimal syntactic modifications to evaluate model robustness.

*Unfold Shorthand Expressions.* This transformation replaces concise instructions with their equivalent long-form representations. For example, the shorthand expression `a += b` can be unfolded into `a = a + b`. This technique has been used in studies such as [61, 62, 63, 64] to analyze the impact of syntactic variations on program behavior.

*Swap Operands.* Swapping operands is a common transformation, which can be either asymmetric or symmetric. In an asymmetric swap, operands of comparison operators are exchanged, such as transforming `a > b` into `b < a`. This has been employed in works like [56, 65, 64]. On the other hand, symmetric swaps involve exchanging operands of commutative operators, such as converting `a + b` to `b + a`, as seen in [61, 66].

Table 5. Categories of transformations, their counts, usage in primary studies, and references.

| Category | No. of transformations | % of primary studies | Used By |
|---|---|---|---|
| Trivial | 19 | 36% | [60, 59, 58, 61, 63, 62, 64, 56, 57, 67, 66, 65, 68, 69, 8, 9] |
| Identifier Renaming | 15 | 86.5% | [70, 60, 59, 71, 72, 73, 74, 75, 58, 76, 65, 77, 78, 79, 80, 63, 81, 82, 83, 84, 85, 86, 87, 39, 64, 88, 57, 89, 90, 91, 56, 69, 68, 67, 61, 92, 8, 9, 66] |
| Data | 16 | 50% | [70, 60, 59, 93, 94, 72, 74, 75, 58, 76, 56, 67, 57, 68, 39, 69, 78, 79, 88, 80] |
| Control Flow | 19 | 58% | [70, 60, 59, 93, 94, 72, 73, 74, 75, 58, 76, 95, 56, 65, 61, 66, 57, 89, 62, 63, 64, 91, 8, 9, 39, 78] |
| Function | 12 | 24.5% | [70, 93, 94, 72, 75, 95, 8, 9, 56, 39, 57] |
| Dead Code Insertion | 27 | 64.5% | [60, 72, 93, 73, 74, 75, 76, 95, 78, 66, 79, 80, 81, 88, 82, 8, 9, 56, 57, 96, 61, 97, 92, 91, 68, 69, 89, 39, 64] |
| API | 3 | 6% | [93, 57, 63] |
| Comment | 3 | 6% | [93, 91, 64, 8] |
| Miscellaneous | 6 | 20% | [93, 71, 75, 76, 8, 9, 57, 67, 64] |
| Combined | 121 | | |

*Remove Unused Code.* This straightforward technique involves removing unused variables, functions, or libraries from the code. Studies by [56, 57, 67] have utilized this transformation to streamline programs and evaluate the impact of dead code on program analysis.

*Split/Aggregate Declarations.* This transformation alters how variables are declared. For example, a split declaration changes `int a, b` into `int a; int b`, while an aggregate declaration performs the reverse operation. This technique has been used in studies such as [56, 62] to assess the effects of declaration styles on code readability and maintainability.

*Split/Aggregate Declaration and Initialization.* This transformation modifies initialization patterns, such as converting `int a = 1` into `int a; a = 1`. This approach has been discussed in works like [62, 64], where it is used to evaluate the impact of initialization patterns on program robustness.

*Exchange Suffix/Prefix.* The exchange of suffix and prefix operators, such as replacing `i++` with `++i`, is another common trivial transformation. This approach has been employed in [65, 62] to assess how subtle syntactic changes affect program behavior without altering its core functionality.

*Add Neutral Element.* Adding a neutral element to logical expressions, such as transforming `true` into `0^1 == 1`, is a transformation explored in [8, 9].

### 4.1.2 Identifier Renaming.
Transformations in this category have been used in 86.5% of studies covered in our literature review. These transformations include renaming code identifiers with various methods to affect or verify model performance. Unlike other surveys, e.g., [51], that classified all identifier renaming methods into a single transformation class, we distinguish these methods based on their unique approaches since the choice of renaming strategy significantly influences model behavior [51].

One of the methods includes the *rename identifier based strategy*, where special techniques are employed to replace identifiers [65, 77, 78, 79, 80, 63, 81, 82, 83, 84, 85]. Another common transformation is *rename identifier with similar words*, which involves renaming identifiers into semantically close words [77, 86, 87]. Techniques such as *rename identifier to template value* (int a → int var1) [88, 57, 89] and *rename identifier to another random identifier* replace an identifier with another random one from the dataset [90, 91, 57], have also been extensively used. Obfuscation techniques, such as *Obfuscate identifier* (int test_counter; → int Bq41wD7l;), aim to replace identifiers with random character sequences [56, 69].

Further examples include *variable rename*, which aligns directly and very well with renaming techniques to preserve functionality while testing robustness [76, 58, 75, 74], and renaming local variables, function parameters, or object fields [73]. State-of-the-art techniques, such as *Semantic-Preserving Identifier Renaming*, apply adversarial perturbations in the embedding space to focus on identifiers like functions or variables, impacting predictions without altering functionality [71]. Other works perform renaming of APIs, arguments, and method names and then analyze how semantic shifts are handled by models [72]. Transformations that rename variables systematically or arbitrarily also play a crucial role in testing models for semantic equivalence [59, 60, 70]. Finally, renaming identifiers with words from a vocabulary [92, 8, 9] and contextual naming methods, such as *Renaming identifier to the context value* (int count → int count_var_1) [88], also emphasize the variety of identifier renaming transformations. In assessing the robustness and generalizability of models under various naming conventions, these instances emphasize the significant role played by identifier renaming.

*4.1.3 Comment Transformations.* Comment-related transformations have been applied in about 9% of the studies presented in our literature review. These code comment modifications include content, position, or structure changes. Comment transformations are practical evaluation techniques used by researchers to test model robustness when analyzing code files containing comments. One of these common transformations is called *delete comment* and involves removing comments entirely from the code as explored in research like [8, 64]. On the contrary, *add comment* involves adding new comments to the code to assess the ability of the model to deal with more annotations [8]. Another strategy is *move comment*, in which case comments are relocated to different locations in the code, thus potentially changing their contextual meaning [8]. More complex transformations involve *move word in comment*, where the rearrangement of particular words in comments is carried out [91], and *delete word in comment*, in which certain words are deleted while preserving the overall structure of the comment [91]. In addition, transformations like *copy word in comment* copy a word within a comment and paste it after the first word to test the model's response to redundancies [91]. Finally, [93] introduces a suite of transformations that insert, delete, or permute comments in the code to test their effect on model predictions. These examples show the varied ways that comment transformations can be used to quantify and analyze model performance and so are central to transformation-based evaluation.

*4.1.4 Data Transformations.* These transformations employed in 45.5% of the primary studies describe changes that modify the representation or manipulation of data in the code. These changes are added to test the adaptability of the model to representational changes in data without changing the behavior of the original program. An example is *use cast expressions*, in which data values are cast to their own type explicitly, i.e., b = (int) a, in which a is an integer variable. While applied in general for type-casting, this transformation applies the property for testing purposes [56, 68]. An example is *Convert string literals to char arrays*, in which string variables are declared via character arrays, challenging the model to comprehend varied string representations [56, 57]. Another example is *convert char literals into ASCII values*, where character literals are substituted with their corresponding ASCII numeric representation (e.g., char c='A' → char c=65), challenging the model on processing equivalent yet differently represented data. *Convert int literals into expressions* involves replacing integer literals with mathematical expressions, such as int b = 8 → int b = 2 * 4, providing alternative representations for numeric values. The other interesting transformation is *convert integers to hexadecimal numbers*, which transforms normal integers into their hexadecimal equivalents (e.g., int b = 48 → int b = 0x30). In addition, *use typed expression* infers data types dynamically according to the keyword typeid, e.g., int b = 0 → typeid (a) b = 0, where a is an integer variable defined previously [56]. Other examples of data transformation are *Def-Use Break*, in which def-use chains are broken to quantify how models recover from variable relation changes and remain correct [70].

Another widely used transformation is *boolean Exchange*, which changes logical values such as true ↔ false, accommodating data representation changes [60, 59, 58, 76]. Such operations as type conversions (e.g., from bool to int) challenge the model to accommodate these conversions [93]. Likewise, the *EncodeStrings* transformation, where literal strings are replaced with calls to functions, shows how data can be encoded differently without changing functionality [94]. Other transformations that fit this category include *plus zero adding* (e.g., a = a + 0) and *return optimal adding* (e.g., return 0 if (1==0) else 1), testing robustness to redundant or conditional expressions [72]. Furthermore, data flow operations are also applied widely, particularly to tasks like Authorship Attribution and Clone Detection, based on variable computations and relationships (e.g., identifier nodes) [74]. There are other experiments that test transformations such as relational operator switching (e.g., a < b → b > a) or inserting equal assignments (e.g., b = -10; → b = b - 10) to evaluate the model adaptability to diverse data manipulations [75]. These examples, as well as the wide variation in data transformations across the literature, highlight the significant role these techniques have in establishing the robustness and flexibility of models with structured changes in data representation.

*4.1.5 Control Flow Transformations.* These transformations are used in 58% of the studies included in this literature review. They are code transformations that alter the code execution process by altering the control flow or control structure of statements. These transformations play an important role in ensuring a model's ability to handle changes in program logic without affecting the program functionality. One example is *convert for-statement to while-statement*, where a for loop is transformed into a while loop and vice versa. For instance, for(init; condition; update;) { body; } is converted into init; while(condition) { body; update; }. This transformation has been explored in [56, 65, 61, 66, 57, 89, 62, 63, 64]. Another transformation, *convert if-else to switch-case*, converts if(condition){body;} into switch(condition){case true: body;} or reverses the process [56, 61, 89, 62, 64]. Transformations like *loop exchange*, which involve replacing one type of loop (e.g., for) with another (e.g., while), are widely used, as demonstrated by [76, 58, 75, 59]. Such transformations, though maintaining the functionality of a program, pose challenges to models to accommodate diverse structural representations.

*Swap independent statements* rearranges the order of two statements that are independent in the control flow graph [61, 91, 89, 64]. Further examples include *Split conditions of if-statements*, which separates logical conditions in if statements (e.g., if(a && b) → if(a) if(b)) [56, 57, 62], and *Convert if-else to conditional expression*, where an if-else block is replaced by a ternary operator (e.g., if(condition){a;} else {b;} → condition ? a : b) [56, 62].

[74] highlights the importance of control flow patterns, such as if, else, for, and while, in the context of successful attacks. Similarly, [73] manipulates structures like if, else, and loops

while ensuring semantic preservation, modifying the program structure but not its functionality. Transformations such as *Wrap expression in if statement* (e.g., wrapping code in `if(true)` or `if(false)`) [8, 9], *Wrap expression in lambda*, which encapsulates expressions in a lambda function [8, 9], and *Swap if-else bodies*, which interchanges the bodies of an `if-else` block while negating the condition [56], further expand the scope of this category.

More advanced techniques, such as those in [94], include *flatten* and *MergeSimple*, which reorganize statements and functions to alter the control flow while preserving semantics. Similarly, [72] demonstrates the use of *Loop Exchange* and other control flow alterations to test model robustness. [59] and [60] emphasize *Loop Exchange* and *Switch to If*, challenging models to adapt to restructured program flows. [70] employs *If-Else Flip*, which modifies the branching structure of an `if-else` statement to assess the model's logic understanding. [95] employs obfuscated conditions to modify program logic without affecting static analysis complexity. These transformations illustrate the diversity and usability of control flow transformations in analyzing the flexibility and resilience of models to complex and heterogeneous program structures, while some other examples support this category even further.

*4.1.6  Function Transformations.* . These transformations have been used in 24.5% of the primary studies, and they are function-level mutations to check model robustness to structural and semantic changes in functions. These changes are meant to check how models react to structural, argument, or definition changes in functions without altering overall program functionality. As one case in point, enhancements such as *Add Function Arguments* add additional parameters to a function using names of the arguments drawn either from an accessible dictionary or inferred programmatically based on the test program, as treated in [8, 9]. Similarly, *Delegate to Function* involves extracting a block of code or expression into a separate function and invoking the new function instead, a technique employed in two different studies by Applis et al. [8, 9]. More complex transformations include *Merge Function Arguments*, where arguments are combined into a struct type and accessed as struct members (e.g., `int add (int a, int b)` → `int add (args ab)`), as explored in [56]. Likewise, *Reorder Function Arguments* changes the sequence of arguments in function calls to assess the model's ability to interpret reordered inputs, as used in [56]. Another example is *Convert Binary Expressions into Functions*, where binary operations like `c = a + b` are extracted into standalone functions, such as `c = add(a, b)`, to evaluate how models handle abstracted computation logic [56].

Recent studies add more dimensions to function transformations. For example, [95] modifies the Function Call Graph (FCG) by adding predetermined functions, explicitly altering the function-level structure while maintaining predictability. Similarly, [72] introduces transformations that add unused arguments or modify function definitions to test how Pre-Trained Models (PTMs) adapt to semantic shifts. Transformations such as *RndArgs, MergeSimple, MergeFlatten, SplitTop, SplitBlock, and SplitRecursive*, as utilized by [94], directly manipulate function arguments and structure, showcasing the versatility of function transformations. Additionally, [93] explores transformations that extract snippets into new functions to evaluate abstraction handling, and [70] tests code completion models with *Independent Swap*, swapping independent statement blocks while ensuring the logical order remains intact. These instances point to the importance of function transformations in assessing the robustness of machine learning models for code analysis and generation tasks.

*4.1.7  API Transformations.* API transformations have been applied in 6% of the studies in our literature review and are the least common form of transformation. These modifications involve altering APIs between equivalent constructs in C and C++ and testing a model's ability to adapt to changes in programming interfaces while maintaining functionality. These transformations target swapping, enabling, or disabling API components that are commonly used in C and C++. One of the simplest transformation techniques in programming is known as *Swap C/C++ Libraries for Reading and Writing*, which replaces standard input/output operations like `printf` from '`<stdio.h>`' with their C++ equivalents, such as `cin/cout` from '`<iostream>`'. This transformation tests how good models can generalize across equivalent but syntactically distinct APIs [57, 62]. Another related transformation is *Use stdin/stdout Instead of Files*, where console input/output is used instead of file-based I/O operations, providing a shift in data handling paradigms [57]. Additionally, transformations like *Enable/Disable Sync Between C/C++ Streams* modify the synchronization behavior between C and C++ streams, challenging models to handle differences in stream handling [57]. Such transformations play a crucial role in evaluating the robustness of models when dealing with program interface level changes across different but equivalent paradigms in C and C++.

*4.1.8 Dead Code Insertion Transformations.* . These transformations have been employed in 64.5% of the studies and represent a very significant category of transformations. These transformations involve the insertion of code that is never executed during the program's execution. The primary purpose is to test the model's ability to handle redundant or irrelevant code constructs without impacting the functional behavior of the program. For instance, [95] employs the "try-catch trap" mechanism to add never-executed function calls, ensuring no logical impact on the program. Likewise, [76] employs Unused Statement, which introduces no-op constructs like unreachable blocks, to challenge model robustness. Another example is [75], which includes dead code insertion techniques such as adding `if (false)` blocks or unused constructs, ensuring semantic equivalence. [74] mines patterns to create redundant constructs, such as `if(false){...}`, testing the model's adaptability. [73] utilizes techniques like adding junk code, including unused variables or inconsequential print statements, to evaluate the model's handling of irrelevant additions. [72] applies techniques such as adding unreachable or unused code, such as `if (1==0): print(0)`, to evaluate the robustness of the model. Moreover, [60] employs methods such as inserting try-catch blocks and unused string declarations to assess the robustness of the model. These studies demonstrate some of the diverse applications of dead code insertion transformations.

One common transformation is *Add Dead Code*, where blocks such as `if(false){...}` are inserted, with the content of the block determined based on a predefined strategy [78, 66, 79, 80, 81]. Another transformation, *Add Print Statement*, involves inserting print statements into the code, even though the output is irrelevant to the program logic [78, 88, 79, 80, 81]. Transformations such as *Add Unused Variable (Strategy)* introduce variables that are never utilized in the program, with their names determined based on a specific strategy [66, 81, 82]. Similarly, *Add Unused Variable (Name from Dictionary)* selects variable names from a predefined dictionary, while *Add Unused Variable (Name from Program Under Test)* sources variable names directly from the program being analyzed [8, 9]. Other examples include *Add Libraries*, which involves inserting unnecessary library imports from other snippets, testing the model's handling of redundant imports [56, 57]. *Transfer Code* takes this further by inserting code from unrelated programs into unreachable parts of the code, such as a dead branch or loop [96, 61]. Additionally, *Insert Unreachable Loop or Branch* places constructs like `if(false)` or `while(false)` blocks into the program to create unexecuted pathways [97, 81]. Finally, *Insert or Delete Empty Statement/Branch/Loop* involves manipulating empty control flow constructs, such as adding or removing empty statements from the code [97, 92]. These transformations highlight the diversity and practical utility of dead code insertion techniques in testing the robustness and adaptability of models. They represent a comprehensive approach to analyzing how models deal with superfluous or unreachable code structures.

*4.1.9 Miscellaneous Transformations.* Around 20% of the studies in our literature review include transformations that do not fit neatly into other defined categories. These transformations are often simple or stylistic but can still significantly impact a model's performance when analyzing or generating code. One example is *Add or Remove Whitespace*, which adjusts the formatting by inserting or deleting whitespace characters without altering the functionality of the code [8]. Another transformation, *Add or Delete Compound Statement*, involves adding or removing curly braces (`{}`) around single-line control flow statements, introducing or eliminating explicit block boundaries [57]. Similarly, transformations like altering formatting or adding curly braces fall into this category, testing the model's robustness against stylistic changes [93]. Transformations like *Add Explicit Return to Main Method* enforce the inclusion of an explicit return statement in the main function, while *Replace Literal Return with Variable Return* changes literal return values to a variable, even if the variable name is not explicitly defined [57]. These transformations test the model's ability to adapt to stylistic variations in code structure. In addition, transformations such as *Statement Permute*, which rearranges and reorders independent statements, further challenge models by introducing structural changes without affecting program semantics [76, 75]. Another notable example is the use of gradient-based virtual perturbations in the embedding space, a novel approach that does not align with conventional transformation types. This method introduces semantic preserving perturbations directly in embedding space, providing unique insights into model robustness [71]. Furthermore, dividing expressions or merging variable declarations has been shown to affect the structure of code while preserving its functionality [75].

Other examples include *Auto-Format Code*, which involves reformatting the code to adhere to a specific style or formatting standard [67], and *Delete Print Statement*, which removes debugging or diagnostic print statements from the code [64]. These transformations test the model's resilience to small stylistic or structural code transformations. These miscellaneous transformations, while less structured than other categories, provide valuable insights into how models handle small, minor, or stylistic modifications that can still affect parsing, analysis, or generation processes. We find that identifier renaming and dead code insertion are the most commonly used transformations. These transformations provide the most freedom to change a code snippet in a way that can be deceiving for a model. Moreover, these transformations allow for features to be transferred from differently labeled code snippets, which could explain why they are so commonly used.

> **Answer to RQ1:** Metamorphic transformations used in model evaluation are very diverse. Among these, *Dead Code Insertion* and *Identifier Renaming* are the most frequently used. The most frequently applied transformations include *Dead Code Insertion* (64.5%), *Identifier Renaming* (86.5%), and *Control Flow* (58%). While *Trivial* (36%), *Data* (50%), and *Function* (24.5%) transformations are widely used for baseline and semantic testing, *API* and *Comment* transformations are the least developed (6% each). The *Miscellaneous* category (20%) encompasses heterogeneous or study-specific transformations.
>
> **Note to Researchers.** Future work should prioritize underdeveloped transformation types such as *API* and *Comment*, and expand beyond the most commonly used techniques to build more comprehensive robustness evaluation frameworks.

Below, we outline several concrete research directions to guide future efforts:

- **Broaden the palette.** API–level rewrites, comment/whitespace edits, and full function-restructuring remain largely unexplored yet mimic many real-world refactoring operations. Studying their impact can uncover blind spots that identifier renaming and dead-code insertion miss.
- **Aim for deeper semantics.** Go beyond surface changes to behavior-preserving transformations that stress data flow, control, and resources (e.g., loop unrolling, memorization), testing if models truly understand program meaning.

- **Quantify transformation quality.** Complement success rates with measures of *naturalness*, compile-rate, and human readability to ensure that generated variants remain realistic and actionable.
- **Cross-language tooling.** Current libraries are Java and Python centric. Developing language-agnostic, AST-level transformation toolkits (covering JavaScript, C#, Go, and UI frameworks) will enable broader benchmarking.
- **Sequence and synergy.** Investigate how combinations of transformations interact. Adaptive schedulers or search algorithms that chain multiple operators may reveal compounding vulnerabilities invisible in single-step tests.
- **Shared artifacts.** Publish open transformation corpora, reference implementations, and automatic validity checkers so future studies can compare on identical perturbation sets and foster cumulative progress.

## 4.2 RQ2: Application Techniques for Metamorphic Transformations

In this section, we analyze the techniques used to apply metamorphic transformations for assessing the robustness of deep code models. We refer to these as *application techniques*, which describe how transformations are **selected, ordered, and executed** during testing. These techniques range in complexity, from simple one-pass methods that apply transformations uniformly or at random, to more advanced approaches such as evolutionary algorithms that optimize transformation combinations, and gradient-based techniques that target model-sensitive features. Some studies perform random dead code insertions without feedback, while others use gradient information or model confidence scores to guide targeted identifier renaming. To improve clarity, we categorize these techniques into distinct families and provide explanations and examples for each. Some studies, e.g., [88], [68], and [39], do not explicitly describe how transformations were applied.

*4.2.1 One-pass Techniques.* One-pass techniques are characterized by the simplicity of their design and the lack of optimization or feedback loops. One-pass techniques use a given set of transformations randomly or systematically without iterative improvement or advanced decision-making processes. For instance, Applis et al. [8] employed a fixed set of transformations to evaluate the robustness of machine learning models in program analysis tasks with a focus on uniform transformations. Similarly, Gao et al. [90] employed a random replacement strategy, replacing identifiers with valid alternatives from a dataset in an unguided and systematic way. Jin et al. [67] employed a set of fixed transformations up to 20, including variable renaming and dead code insertion, to generate augmented data sets for contrastive pre-training. This strategy favors simplicity in that the transformations are applied without optimization or adaptive feedback. Liu et al. [91] also followed a one-pass strategy by stochastically applying a fixed set of transformations, such as variable renaming and dead-code insertion, to generate semantically equivalent code variants. Rabin et al. [89] systematically applied transformations across either a single location or all eligible locations in the program, relying solely on predefined rules rather than adaptive mechanisms. Extending these approaches, Li et al. [95] implemented a strategy involving straightforward transformations like inserting dead code and adding valueless calls, applying them in an unguided manner typical of one-pass techniques. Pei et al. [76] similarly employed simple techniques, using variable renaming and statement permutations without feedback or optimization. Springer et al. [58] also followed a fixed strategy, applying loop exchanges and variable renaming in a consistent, predefined manner. Le-Cong et al. [75] adopted a strategy involving systematic application of predefined transformations such as renaming variables, adding unused statements, and swapping statements (e.g., `if(false)`), without any iterative optimization. Nguyen et al. [74] employed variable renaming as part of their strategy, focusing on simplicity. Zhang et al. [73] followed a similar unguided strategy by using random insertions of dead code and straightforward renaming tasks. Li et al. [94, 93] developed techniques based on predefined transformations like renaming variables and adding unused statements, systematically applying them without any optimization mechanisms. Rabin et al. [59, 60] demonstrated trivial techniques by systematically applying simple transformations such as Boolean value swapping and loop exchanges across eligible locations. Hooda et al. [70] adopted a fixed strategy of applying semantic-preserving mutations, such as flipping if-else statements or randomizing variable names, maintaining the predefined and unguided nature characteristic of trivial techniques. These sampling-based techniques focus on simplicity, consistency, and ease of application. Although adequate for basic robustness assessments, they lack the adaptability and effectiveness offered by more advanced methods based on optimization or iterative feedback.

*4.2.2 Evolutionary Techniques.* Evolutionary techniques take advantage of evolutionary algorithms to determine which transformation to apply based on the optimization of a fitness function. For example, Applis et al. [9] use a genetic algorithm to find successful transformation sets. The fitness function in this approach maximizes the drop in the F1 score, with individuals represented as sequences of transformations applied to the abstract syntax tree. Similarly, Ferretti and Saletta [96] use Grammatical Evolution, a genetic algorithm, to inject code snippets into given examples. The confidence of the model acts as the fitness function, guiding the evolutionary process, though the details of the crossover and mutation operators are not specified. Li et al. [95] introduce the Apoem framework that makes use of a multipopulation co-evolution algorithm to construct adversarial changes. This approach enhances performance by taking various levels of granularity into account and implementing evolutionary mechanisms to generate varied and robust adversarial examples. Yang et al. [85] take a hybrid approach, initially applying a greedy search to identify the most important identifiers by replacing them with non-informative tokens and measuring the change in confidence in the model. If no prediction change occurs, a genetic algorithm is employed to find the optimal combination of variable replacements. Here, individuals are encoded as sets of pairs of original variables and their substitutions, with model confidence serving as the fitness function. Similarly, [84] employs masking methods to detect important identifiers by substituting them with tokens such as <UNK> and assessing the causal effect on model confidence. From this knowledge, they produce an initial population of perturbed identifiers and apply the NSGA-II algorithm [98], optimizing adversarial loss, semantic similarity, and modification rate. These methods illustrate the ability and strength of evolutionary algorithms for designing effective transformation pipelines driven by clearly defined objectives and fitness functions.

*4.2.3 Gradient-Based Techniques.* Gradient-based techniques depend on computing the gradient with respect to the input to determine which transformations to use. Yefet et al. [82] utilize the gradient to calculate the loss concerning identifiers when testing a sample. They then identify the token that causes the maximum disruption by conducting a local search in the embedding space. Building upon this method, Chen et al. [65] calculate a Jacobian matrix to pinpoint the most important tokens and apply the gradient-based approach by Yefet et al. to replace identifiers. Structural transformations are targeted to affect these critical tokens the most. Henkel et al. [78] take a slightly different approach, randomly sampling sequences of transformations (e.g., 10 sequences of 5 transformations) and then using a gradient-based method inspired by Yefet et al. to resolve the inconsistencies or "holes" created by the transformations. Meanwhile, Liu et al. [56] train a surrogate model to approximate the victim classifier and use the Jacobian matrix to identify the most significant features, such as identifiers. They iteratively select transformations that maximize the cumulative feature importance, applying the most impactful transformations until the model's prediction changes. Li et al. [71] use a gradient-based adversarial training paradigm directly, using Projected Gradient Descent (PGD) to calculate perturbations in the embedding space. The method is designed to generate adversarial examples in a systematic manner for evaluating model robustness. Yang et al. [72] use gradient-based methods, such as adversarial training, to identify significant features and manipulate them in a systematic manner, to enhance model robustness through the identification of the most sensitive features. Nguyen et al. [74] incorporate gradient-guided transformations in their framework, CARROT. These include techniques like gradient-based identifier renaming and dead-code insertion. However, in the study's black-box setting, these methods are not directly employed, highlighting a limitation of gradient-based techniques in scenarios without access to model gradients.

*4.2.4 Sampling-Based Techniques.* Sampling-based techniques determine identifier replacements by sampling from a distribution, often guided by specific selection criteria. Zhang et al. [83] introduced the Metropolis-Hastings Monte Carlo (MHM) method, where new identifiers are stochastically selected from a vocabulary and accepted with a probability proportional to the decrease in the model's confidence. This approach ensures that replacements disrupt the model's predictions effectively. Building upon MHM, Ding et al. [77] create code variants by replacing identifiers but constraining replacements to semantically similar words, refining the method to focus on meaningful substitutions. Zhang et al. [92] extend MHM further by using gradient information in the embedding space during new identifier selection. This gradient-guided sampling enables more specific substitutions that take advantage of the model's weaknesses. They also use structural transformations randomly for the sake of adding diversity among the generated variants of the code. Nguyen et al. [74] employ a sampling-based strategy by choosing attack positions and patterns according to impact estimation using statistical scores. This strategy ensures that the sampling process biases impactful transformations without compromising the randomness in the selection process. Le-Cong et al. [75] employ a sampling-based strategy for certain transformations like renaming variables. In their method, replacements are sampled at random from a distribution, with decisions based on pre-trained models like CodeBERT (e.g., RenameVariable-2). The random sampling adds a degree of randomness while maintaining semantic correctness, thus enhancing the robustness testing by embracing a diverse set of possible transformations. These methods apply probabilistic sampling in addition to other restrictions, e.g., semantic similarity and gradient information, to systematically examine identifier replacements and test the model's robustness.

*4.2.5 Transfer-Based Techniques.* Transfer-based approaches attempt to deceive deep code models by transferring features from a differently labeled code snippet into the target one. Gao et al. [61] employ transformations that reduce the model's prediction probability standard deviation, thus indicating the most significant features of a code snippet. They then mislead the model by inserting significant features from other programs as dead code. Similarly, Na et al. [97] identify susceptible positions in the code by adding uninformative tokens and measuring the resulting decrease in model confidence. They then add salient features from unrelated programs to these critical positions with the goal of outsmarting the model's understanding. Tian et al. [63], on the other hand, use a different approach by finding reference inputs from the second most probable class as predicted by the model. They subsequently employ transformations that cause the attacked input to mirror the reference input, exploiting the similarities to mislead the model. These methods skew the transfer of features between snippets of code, obscuring the perception of the model and rendering them highly effective at assessing model resilience to adversarial attacks.

*4.2.6 Embedding-Based Techniques.* Embedding-based techniques try to maximize the distance between original and rewritten code snippets in the embedding space, subtly transforming the representation of the program without altering its semantics. Jia et al. [80] merge techniques from Henkel et al. [78] and Srikant et al. [79] to implement transformations to maximize the discrepancy between input programs in embedding space. This method attempts to destabilize the model's internal representation of the code by intentionally changing its properties. Springer et al. [58] use token embeddings, where they use metrics like the L2 norm for adversarial token replacement. This method manipulates the embedding space directly in an attempt to evaluate the model's robustness against token-level perturbations. Likewise, Pei et al. [76] use Aut(PDG)-equivariant self-attention layers such that transformations are semantically valid and investigate embedding-space representations to analyze model robustness. Le-Cong et al. [75] use semantic-preserving transformations such as variable substitution with pre-trained models such as CodeBERT (e.g., RenameVariable-2). These approaches attack the embedding space to preserve program equivalence but disrupt the model's internal representations. Li et al. [71] also deal with embedding-based methods, which target programming language model embeddings to be semantically preserved and tested for robustness through systematic perturbations. Yang et al. [81] take this even further by employing a genetic algorithm to find names that are optimally different in the embedding space, confirmed using a surrogate encoder. Moreover, they embed code snippets in prominent positions, identified by introducing an explanatory token and measuring the resulting decrease in model confidence. They check for an increase in loss produced by these insertions, leading to further mismatch in the model's embeddings. These methods effectively leverage embedding space manipulations to test the strength and flexibility of deep code models, uncovering their weaknesses in structure-preserving semantic transformations.

*4.2.7 Miscellaneous Techniques.* Miscellaneous techniques include a wide variety of techniques that do not fall under the above categories. Quiring et al. [57] use Monte Carlo Tree Search [99] to iteratively choose transformations most likely to change a model's prediction. Srikant et al. [79] use alternating optimization to find the most impactful positions and transformations to add. Zhang et al. [86] use a method to identify vulnerable identifiers by replacing them with uninformative tokens and evaluating the model's confidence. They also fine-tune their replacements with CodeBERT, generating contextually relevant replacements when required. Zhang et al. [64] compared multiple algorithms, such as random search, a genetic algorithm for edit distance maximization, a Markov Chain Monte Carlo algorithm for maximization of perplexity scores, and a deep reinforcement learning algorithm with dissimilarity and clone detection performance rewards. Notably, none of these methods make use of model feedback to direct the transformations. Zhou et al. [87] target the cosine distance between identifiers and program embeddings to detect and substitute important identifiers. Liu and Zhang [66] design structural mutations to produce various code snippets, which are passed through a trained network to find those most susceptible to being misclassified. They take it one step further by reinforcing these snippets, filling in gaps to make misclassification more likely. Tian et al. [62] formulate adversarial sample generation as a Markov decision process and use Q-learning with rewards given by model confidence decreases. Wei et al. [69] focus on transformations that excite the most new neurons, trying the code space thoroughly.

---

**Answer to RQ2:** Metamorphic testing techniques vary significantly in complexity and application. The most common techniques are *one-pass techniques*, which apply transformation without optimization or a feedback loop. More sophisticated approaches use optimization methods, such as *evolutionary algorithms*, to refine transformations iteratively and *gradient-based* techniques. Other techniques, such as *sampling-based*, *embedding-based*, *transfer-based*, and *miscellaneous*, are less frequently applied.
**Note to Researchers.** Future work should aim to systematically compare these techniques and assess their trade-offs in terms of efficiency, effectiveness, and generalizability.

---

Below, we outline several concrete research directions to guide future studies:

- **Benchmark systematically.** Establish common, publicly available benchmarks (models, tasks, and datasets) and report query budget, run-time, and success criteria so that one-pass, gradient-based, evolutionary, and other techniques can be compared on equal footing.
- **Develop hybrid pipelines.** Combine complementary ideas e.g., use gradient saliency to seed an evolutionary search, or integrate embedding-distance objectives into genetic fitness to push beyond the ceiling of any single technique.
- **Embrace black-box constraints.** Many industrial LLMs expose only logits or final predictions. Explore efficient surrogate model and transfer-based strategies to make MT practical when gradients are unavailable.
- **Leverage static & dynamic analysis.** Traditional program analysis cues (def-use chains, control dependence, taint flow) can guide transformation choice and sharply reduce the search space.
- **Track efficiency, not just effectiveness.** Report the number of queries, tokens mutated, and wall-clock time per successful test; real-world pipelines must balance power with cost.
- **Release tooling.** Open-source generators, fitness functions, and evaluation harnesses will accelerate replication and drive *de facto* standards for MT on deep code models.
- **Study generalization.** Test whether MT techniques transfer across languages, tasks, or model families—robustness without generalization offers limited assurance.

### 4.3 RQ3: Investigated Programming-related Tasks

Table 6 presents the downstream tasks that the models under test perform. The supplementary material includes the metrics for evaluating model robustness with these tasks, along with metrics for comparing model input.

We observe that the most commonly tested tasks are clone detection, method name prediction, and authorship attribution, accounting for 29%, 29%, and 18% of the literature, respectively. This prevalence is likely due to the availability of widely-known benchmark datasets, such as BigCloneBench [100] for clone detection, the datasets for training code2vec [29] for method name prediction, and the Online Judge dataset [101] for authorship attribution. Clone detection involves identifying duplicate or similar pieces of code, which is critical for maintaining code quality and identifying redundancies [74]. Method name prediction focuses on predicting descriptive names for functions or methods based on their content, aiding code readability and maintenance [60]. Authorship attribution, which determines the author of a code snippet, has applications in plagiarism detection and software forensics [74].

Among other important tasks, 18% is taken up by code summarization and 11% by functionality classification. Code summarization entails generating natural language summaries of functions in codes, an important part of documentation and for an understanding of complex codes [93]. Functionality classification categorizes code snippets based on their functionality, aiding in understanding and organizing large codebases [66]. Defect detection and vulnerability detection each represent important areas, representing 13.5%. Defect detection is focused specifically on identifying code bugs in a quest for software reliability [71], and vulnerability detection seeks to reveal code security vulnerabilities, and in doing so, helps maintain software integrity and reduces vulnerability to exploitation [94].

Tasks such as code completion (9%), code translation (6.5%), and code repair (4.5%) have received attention in studies, even with fewer studies focused specifically on them. Code completion predicts the next code token or statement, assisting developers during software development [70]. Code translation converts code from one programming language to another, aiding in software interoperability [72]. Code repair automatically fixes bugs in the code, often leveraging model-generated fixes to improve code quality [75]. Similarly, comment generation (2%) creates explanatory comments for code snippets, facilitating better documentation [87], while type inference (2%) predicts type annotations for variables or functions, aiding developers in maintaining and refactoring code [39]. Code search (6.5%) involves finding relevant code snippets in large repositories, facilitating code reuse, and improving development efficiency [93].

Table 6. Code tasks that the model under test is performing.

| Task | Tested by | Fraction of studies |
|------|-----------|---------------------|
| Clone detection | [74, 80, 66, 92, 86, 64, 97, 63, 84, 81, 85, 67, 91] | 29% |
| Method name prediction | [60, 59, 58, 76, 9, 65, 90, 78, 67, 66, 89, 79, 82] | 29% |
| Authorship attribution | [74, 97, 57, 63, 81, 85, 84, 56] | 18% |
| Code summarization | [93, 73, 80, 69, 8, 77, 66, 84] | 18% |
| Functionality classification | [66, 62, 63, 83, 82] | 11% |
| Defect detection | [71, 76, 97, 63, 81, 84] | 13.5% |
| Vulnerability detection | [94, 74, 96, 92, 63, 85] | 13.5% |
| Code completion | [70, 93, 80, 88] | 9% |
| Code translation | [72, 91, 84] | 6.5% |
| Code repair | [75, 68] | 4.5% |
| Comment generation | [87] | 2% |
| Type inference | [39] | 2% |
| Code search | [93, 71, 91] | 6.5% |
| Malware Detection | [95] | 2% |
| Function Similarity Detection | [76] | 2% |
| Function Signature Prediction | [76] | 2% |
| Memory Region Prediction | [76] | 3% |
| Code Question Answering | [71] | 2% |

Several niche tasks are also studied but to a much lesser extent, each contributing approximately 2–3% of the literature. Malware detection identifies malicious code within programs, enhancing software security [95]. Function similarity detection identifies semantically equivalent functions for optimization or refactoring [76]. Function signature prediction predicts the signatures of functions, such as input and output types, to enhance API usability [76]. Memory region prediction forecasts specific memory regions accessed by programs, useful in low-level system optimizations [76]. Lastly, code question answering involves responding to questions about code snippets, helping with debugging, and explaining complex logic [71].

> **Answer to RQ3:** The most commonly tested tasks in metamorphic robustness evaluations are Clone Detection (29%), Method Name Prediction (29%), and Authorship Attribution (18%). The popularity of these tasks is likely due to the availability of well-established benchmark datasets (e.g., BigCloneBench for clone detection).
>
> **Note to Researchers**. The focus on a few dominant tasks suggests that metamorphic testing for deep code models lacks diversity in task coverage. Expanding evaluations to underrepresented (e.g., code repair, malware detection, memory prediction) or completely new tasks (like code generation and test case generation) could reveal new robustness vulnerabilities and improve generalization assessments.

Below, we outline several concrete research directions to support future work:

Table 7. Top models under test with more than one study in primary literature.

| Model | Used by | N. of Papers |
|---|---|---|
| CodeBERT [32] | [71, 74, 72, 8, 61, 90, 66, 91, 97, 63, 81, 85, 92, 86, 84] | 15 |
| LSTM [102] | [94, 39, 67, 57, 63, 92, 83, 86, 64, 90] | 10 |
| GraphCodeBERT [103] | [71, 74, 72, 76, 66, 91, 97, 63, 81, 85, 86, 84] | 12 |
| Seq2seq [104] | [73, 93, 94, 65, 77, 78, 80, 89, 79, 87, 90] | 11 |
| Code2vec [29] | [60, 59, 76, 9, 61, 67, 66, 89, 69, 82] | 10 |
| Code2seq [30] | [60, 58, 76, 78, 67, 66, 79, 69] | 8 |
| CodeT5 [33] | [72, 76, 66, 97, 81, 84] | 6 |
| GGNN [105] | [76, 39, 61, 89, 82] | 5 |
| ASTNN [106] | [90, 62, 92, 83, 64] | 5 |
| GPT-Based [35] | [73, 76, 72, 66, 88] | 5 |
| TBCNN [101] | [90, 92] | 2 |
| PLBART [107] | [72, 66] | 2 |
| GCN [108] | [95, 39] | 2 |
| SequenceR [109] | [75, 68] | 2 |
| Recoder [110] | [75, 68] | 2 |
| CodeLlama [111] | [76, 73] | 2 |

- **Broaden task coverage.** Extend MT studies to underrepresented yet high value tasks, e.g., *code generation, program repair, test case synthesis, performance optimization*, and *malware/vulnerability mitigation*. These scenarios exercise different reasoning capabilities and may surface new robustness gaps.
- **Security critical evaluations.** Robustness under metamorphic transformations should be assessed on tasks with tangible security impact (defect, vulnerability, and malware detection), where false negatives carry real world risk.
- **End-to-end pipelines.** Move beyond single step tasks: evaluate MT effects across development workflows (e.g., *completion → compile → unit-test*) to capture error propagation.
- **Multitask, multilingual language suites.** Publish shared benchmark collections that pair each task with language diverse, transformation ready datasets; this will enable systematic cross-task comparisons of robustness.
- **Task-specific metamorphic relations.** Design transformation families that reflect the semantics of emerging tasks, for instance, API-level rewrites for *code translation* or logical post condition permutations for *specification-driven synthesis*.

### 4.4 RQ4: Models Under Test

A wide variety of models have been tested for robustness using metamorphic testing, demonstrating the diversity of approaches to evaluate deep-code models. Table 7 highlights the most commonly used models in primary studies, along with additional models that showcase the breadth of tools researchers employ in this field.

*CodeBERT* [32], used in 15 studies, is widely tested for tasks such as code summarization, clone detection, and defect detection. Researchers, including Applis et al. [8], Gao et al. [61], and Zhou et al. [84], leverage its pre-trained contextual embeddings to evaluate robustness under transformations like identifier renaming and dead-code insertion.

*LSTM-based models* [102], appearing in 10 studies, are primarily used for sequence-related tasks such as method name prediction and code completion. Zhang et al. [86] and Jain et al. [67] tested these models in adversarial settings to analyze their ability to handle structural changes in code.

Table 8. Overview of prominent models tested in the literature, categorized by type, architecture, and tasks.

| Type | Architecture | Model | Related SE Tasks | Usage (%) |
|------|-------------|-------|------------------|-----------|
| Understanding | Encoder-only | CodeBERT [32] | Clone detection, Defect detection, Vulnerability detection, Method name prediction, Functionality classification | 62.2 |
| | | GraphCodeBERT [103] | Clone detection, Defect detection, Vulnerability detection, Code summarization | |
| | | LSTM [102] | Authorship attribution, Clone detection, Method name prediction | |
| | | GGNN [105] | Functionality classification, Vulnerability detection | |
| | | Code2vec [29] | Method name prediction, Defect detection | |
| Understanding & Generation | Encoder-decoder | CodeT5 [33] | Code summarization, Code completion, Code translation, Code repair | 42.2 |
| | | Seq2seq [104] | Code summarization, Code translation, Comment generation | |
| | | PLBART [107] | Code summarization, Code repair | |
| | | Code2seq [30] | Code summarization, Code captioning | |
| | | GCN [108] | Defect detection | |
| Generation | Decoder-only | GPT-based [35] | Code generation, Test case generation, Code completion | 26.7 |
| | | Recoder [110] | Code repair | |
| | | SequenceR [109] | Code repair | |
| | | ASTNN [106] | Code classification, Defect prediction | |
| | | CodeLlama [111] | Code generation, Completion | |

*GraphCodeBERT* [103], appearing in 12 studies, integrates graph representations to capture the structural information of code. Studies by Liu et al. [66] and Yang et al. [81] employ this model for evaluating transformations such as node swapping and function argument modification.

*Seq2seq models* [104], used in 11 studies, are predominantly applied to translation and summarization tasks. These models are tested by researchers like Chen et al. [65] and Jia et al. [80] to evaluate their robustness against transformations such as control flow changes and token substitutions.

*Code2vec* [29], used in 10 studies, is particularly notable for its embedding-based approach to method name prediction and function-level transformations. Gao et al. [61] and Yefet et al. [82] apply this model to test identifier-level manipulations.

In addition to these top models, several others are noteworthy. *Code2seq* [30], used in 8 studies, is applied to summarization and sequence-level tasks, as seen in works like Wei et al. [69]. *CodeT5* [33], appearing in 6 studies, leverages a text-to-text paradigm for diverse transformations, such as those tested by Na et al. [97]. Graph-based models such as *GGNN* [105] (5 studies) and *ASTNN* [106] (5 studies) are also widely tested, showcasing their utility in tasks like defect detection and structural analysis.

*GPT-Based models* [35], used in 5 studies, are employed for various natural language and code generation tasks, such as those evaluated by Zhang et al. [73]. Models like *TBCNN* [101], *PLBART* [107], *GCN* [108], *SequenceR* [109], and *Recoder* [110] appear less frequently, each used in 2 studies. These models are often applied to niche tasks like vulnerability detection, repair generation, and structural analysis. Models like CodeBERT and GraphCodeBERT receive extensive attention due to their versatility, while others cater to specific tasks or innovative robustness testing methodologies.

Considering the work of Hou et al. [53] and building on our findings from Tables 7 and 6, Table 8 categorizes models into three types, *Understanding*, *Understanding & Generation*, and *Generation* based on their architectural design and alignment with software engineering (SE) tasks. Encoder-only models dominate the *Understanding* category, accounting for 62.2% of the unique studies reviewed. Notable examples include CodeBERT and GraphCodeBERT, which excel in tasks such as clone detection, defect detection, vulnerability detection, and method name prediction. These tasks emphasize the models' ability to process and comprehend source code.

The *Understanding & Generation* category consists of encoder-decoder models, representing 42.2% of the studies. Models such as CodeT5 and Seq2seq perform tasks that require both comprehension and generative capabilities. These models are particularly effective in tasks like code summarization, code translation, and code repair, reflecting their versatility in addressing real-world SE challenges. Decoder-only architectures fall under the *Generation* category, comprising 26.7% of the studies. Models such as GPT-based architectures are prominent in tasks like code generation, test case generation, and code completion, which are critical for advancing automated software engineering.

Narrowly focused models, including Recorder and SequenceR, specialize in generative tasks like code repair. By integrating insights from prior literature [53], and our systematic review, this categorization highlights not only the current state of model capabilities but also critical gaps. For example, tasks such as test case generation and comment generation remain underexplored, particularly in the context of robustness evaluations. Addressing these gaps is essential to align academic advancements with the practical needs of software engineering, ensuring comprehensive and robust model evaluations.

---

**Answer to RQ4.** A diverse set of models has been tested for robustness using metamorphic testing. The most frequently evaluated models are CodeBERT (15 studies), GraphCodeBERT (12 studies), Seq2seq models (11 studies), Code2vec (10 studies), and LSTM-based models (10 studies). W.r.t model architectures, encoder-only models such as CodeBERT are most prevalent (62.2%), which excel in defect detection and functionality classification tasks. Encoder-decoder models (42.2%), such as CodeT5, strike a balance by facilitating both understanding and generation tasks, whereas decoder-only models (26.7%), such as GPT-based models, are underutilized, particularly for test case generation and comment generation.
**Note to Researchers.** The focus on encoder-only models suggests a significant gap in evaluating generative architectures. Robustness assessments should expand to include decoder-based and encoder-decoder models, especially for generation-oriented tasks.

---

To support this shift, we propose the following concrete directions for future research:

- **Cover more architectures.** Extend metamorphic robustness studies to include encoder-decoder models (e.g., CodeT5) and decoder-only models (e.g., CodeLlama, StarCoder) to better reflect real-world tool usage.
- **Add recent open models.** Evaluate the robustness of newly released open-source models such as *Codestral*, *DeepSeek-Coder*, *Magicoder*, and *Phi-2*, which are notably absent in current robustness benchmarks.
- **Stress generative abilities.** Design metamorphic transformations tailored to code generation, test case synthesis, and long-form repair to assess generative model behavior under perturbation.
- **Compare across languages.** Use identical architectures across multiple programming languages (e.g., Java, Python, C) to isolate model-specific versus language-specific robustness issues.

Table 9. Downstream languages used for robustness evaluations, along with the number of papers.

| Language | Used by | N. of Papers |
|---|---|---|
| Java | [60, 59, 93, 71, 72, 74, 75, 58, 76, 95, 9, 8, 65, 77, 61, 68, 90, 78, 80, 66, 91, 97, 89, 79, 63, 69, 81, 85, 82, 92, 86, 84, 87] | 33 |
| Python | [70, 93, 71, 72, 73, 74, 58, 90, 78, 80, 88, 91, 97, 79, 63, 81, 85, 84, 87] | 19 |
| C | [94, 74, 96, 90, 97, 57, 62, 63, 81, 85, 92, 83, 86, 64] | 14 |
| C++ | [56, 97, 57, 62, 63, 81, 92, 83, 86, 64] | 10 |
| Javascript | [71, 39, 67, 91, 84] | 5 |
| C# | [96, 90, 82, 84] | 4 |
| Go | [71, 90, 91, 84] | 4 |
| Ruby | [71, 90, 91, 84] | 4 |
| PHP | [71, 90, 91] | 3 |
| Typescript | [39] | 1 |

- **Release shared artifacts.** Share transformation scripts, robustness-modified checkpoints, and benchmark tables to enable reproducibility and cumulative progress in metamorphic testing for code.

## 4.5 RQ5: Programming Languages

Table 9 summarizes the statistics related to the programming languages evaluated in the related studies on metamorphic testing for deep code models. Java, Python, and C/C++ remain the most commonly tested languages, reflecting their widespread use in programming and the availability of datasets. These languages dominate robustness evaluations due to their prevalence in software development and their utility in various application domains.

Java is the most frequently evaluated language, appearing in 33 studies. It is commonly used for tasks like clone detection, defect detection, and method name prediction, as seen in studies such as Applis et al. [9], Jia et al. [80], and Zhou et al. [84]. The popularity of Java is supported by its mature tooling ecosystem and well-established datasets.

Python, evaluated in 19 studies, is widely used for code summarization and translation tasks. Studies such as Na et al. [97] and Zhou et al. [84] leverage Python's flexibility and simplicity to evaluate robustness under various transformations.

C, appearing in 14 studies, and C++, used in 10 studies, are primarily tested for tasks like vulnerability detection and code optimization. Their inclusion reflects their importance in systems programming and the availability of challenging robustness tasks. Key studies include Zhang et al. [86] and Tian et al. [63].

Other languages, though less frequently tested, provide valuable insights. JavaScript, evaluated in 5 studies, is primarily tested for web development-related tasks, as demonstrated by Bielik et al. [39]. C#, appearing in 4 studies, and Go, also evaluated in 4 studies, reflect their growing adoption in enterprise and cloud-native applications. Additionally, Ruby (4 studies), PHP (3 studies), and TypeScript (1 study) illustrate the diversity of languages tested, catering to specific use cases and developer communities. The variety of languages tested highlights the flexibility of robustness evaluation frameworks and the need for generalizability in model performance across different programming paradigms.

> **Answer to RQ5**. In terms of programming languages, the majority of the literature focuses on Java (33 papers) and Python (19 papers), languages such as JavaScript (5 papers), C# (4 papers) and PHP (3 papers) being grossly underrepresented. This deviation from current industry practice, especially where JavaScript and C# reign supreme in tools like GitHub Copilot, is a critical need for research focus.
> **Note to Researchers.** The field needs broader language coverage that aligns with industry practice.

To support this direction, future work could explore the following research opportunities:

- **Mind the practice gap.** Modern tooling such as GitHub Copilot is heavily used for JavaScript, C#, and PHP. Future metamorphic evaluations should mirror this reality.
- **Cover more paradigms.** Bring in functional (e.g., Haskell, OCaml), system level (Rust), and scripting (Bash) languages to test whether findings generalise across language families.
- **Design cross-language relations.** Propose metamorphic relations that hold regardless of source language, such as invariance under comment removal or identifier renaming, to enable head to head model comparisons.
- **Share language modules.** Release reusable parsers and transformation plugins so the community can add a new language without rewriting infrastructure.
- **Report language strata.** Always break down robustness results by language; this flags hidden weaknesses that averaged scores can mask.

### 4.6 RQ6: Employed Datasets

The datasets used for evaluating robustness are summarized in Table 10. Among these, the `CodeSearchNet` dataset [112] stands out as the most commonly used, appearing in 12 studies. Other frequently employed datasets include `Online Judge` [101], `BigCloneBench` [100], `java-small` [29], and `Devign` [106]. These datasets provide a diverse range of benchmarks that allow researchers to evaluate the robustness of models across various tasks and transformations.

`CodeSearchNet` [112], featured in 12 studies, is widely used for tasks such as code search and summarization. Researchers like Gao et al. [61] and Chen et al. [65] have leveraged this dataset to test robustness under transformations such as identifier renaming and structural changes, making it a go-to resource for evaluating semantic-preserving methods.

`Online Judge` [101], cited in 6 studies, is a valuable dataset for clone detection and function similarity tasks. Studies such as those by Zhang et al. [86] and Tian et al. [63] have employed this dataset to assess the impact of control flow transformations and token-level manipulations on model robustness.

`BigCloneBench` [100], appearing in 5 studies, is the benchmark of choice for clone detection due to its extensive collection of functionally similar code pairs. Yang et al. [81] and Liu et al. [66] have used this dataset to evaluate how well models handle subtle semantic-preserving transformations.

`java-small` [29], used in 7 studies, focuses on tasks such as method name prediction and code summarization. Applis et al. [9] and Henkel et al. [78] have applied this dataset to evaluate how structural changes, such as node reordering or statement swapping, impact model performance.

`Devign` [106], cited in 5 studies, is widely used for defect detection. Gao et al. [90], and Na et al. [97] have tested models on this dataset using transformations like dead-code insertion and loop unrolling, which simulate realistic coding errors.

Beyond these, other notable datasets also play critical roles in robustness evaluations. `java-large` [29], cited in 6 studies, is particularly significant for large-scale tasks such as method name prediction and summarization. This dataset provides a more comprehensive benchmark than its smaller counterpart, *java-small*, and has been utilized by studies such as Srikant et al. [79] and Yefet et al. [82]to evaluate the robustness of the modelthe model against larger and more complex codebases.

Table 10. Datasets with at least two citations in primary studies, along with the number of papers using them and their primary purpose.

| Dataset | Used by | #Papers | Purpose |
|---|---|---|---|
| CodeSearchNet [112] | [93, 71, 8, 61, 90, 78, 67, 88, 66, 91, 80, 65] | 12 | Benchmark for code retrieval and search tasks. |
| Online Judge [101] | [90, 62, 63, 92, 83, 86] | 6 | Dataset for solving algorithmic programming problems. |
| BigCloneBench [100] | [66, 97, 63, 81, 85] | 5 | Code clone detection and similarity analysis. |
| java-small [29] | [60, 58, 9, 65, 78, 89, 69] | 7 | Code classification and function summarization. |
| Devign dataset [106] | [90, 97, 63, 85, 81] | 5 | Dataset for vulnerability detection in code. |
| java-large [29] | [60, 58, 61, 89, 79, 82] | 6 | Benchmark for code classification and modeling. |
| Py150k [113] | [73, 58, 78, 80, 79] | 5 | Dataset for Python program modeling and code completion. |
| Google Code Jam dataset [114] | [97, 63, 85] | 3 | Programming competition dataset for problem solving. |
| Online Judge Clone [115] | [92, 86, 64] | 3 | Dataset for code clone detection in competitive programming tasks. |
| java-med [29] | [60, 58, 66, 89] | 4 | Dataset for Java function classification and summarization. |
| CodeChef [116] | [63, 92] | 2 | Dataset for algorithmic programming competition problems. |
| Java dataset by [117] | [77, 87] | 2 | Dataset for Java function summarization tasks. |
| Defects4J [118] | [71, 75, 68] | 3 | Bug-fix and fault localization benchmark for Java programs. |
| Google Code Jam compiled [119] | [56, 57, 81] | 3 | Dataset for code generation from competitive programming problems. |
| Ruby dataset [71] | [91, 84, 90] | 3 | Dataset for Ruby function classification and summarization. |

Py150k [113], appearing in 5 studies, is a Python-specific dataset often employed for tasks like code summarization and translation. Studies like Jia et al. [80] and Zhang et al. [73] have leveraged this dataset to assess how well models handle Python's unique syntax and semantic variations under adversarial transformations.

*java-med* [29], used in 4 studies, bridges the gap between the small-scale *java-small* and large-scale *java-large* datasets. It is commonly applied for medium-complexity tasks, as seen in studies like Rabin et al. [89] and Springer et al. [58], where it has been used to evaluate structural and token-level transformations.

Additionally, datasets like Google Code Jam [114] (3 studies), and Online Judge Clone [115] (3 studies) focus on competitive programming and clone detection, respectively. Defects4J [118], appearing in 3 studies, targets defect detection, providing a robust benchmark for testing models' ability to identify and fix bugs. Specialized datasets like Ruby dataset [71] and CodeChef [116] reflect the need for language-specific evaluations.

The variety of datasets pinpoints the importance of standardized benchmarks in ensuring fair and consistent comparisons across studies. Researchers can comprehensively evaluate robustness and identify potential limitations in their models by testing on diverse datasets like java-small, Py150k, and Devign.

---

**Answer to RQ6:** Robustness evaluations rely heavily on a few key datasets, with `CodeSearchNet`, `Online Judge`, `BigCloneBench`, `java-small`, and `Devign` among the most frequently used. These datasets primarily support tasks like clone detection, defect detection, method name prediction, and code summarization. Other notable datasets include `java-large`, `Py150k`, `java-med`, `Google Code Jam`, and `Defects4J`, reflecting the diversity of tasks and programming languages.

**Note to Researchers**: Despite the variety of datasets, robustness evaluations remain concentrated on a few widely-used benchmarks. Expanding evaluations to less frequently used or new datasets and language-specific corpora could improve the generalizability of findings and provide deeper insights into model vulnerabilities. This is particularly critical w.r.t. possible data leakage (see Section 5 and overfitting towards widely used datasets).

---

We recommend the following future research directions:

- **Move beyond the "big five".** Incorporate under-used corpora (e.g., HumanEval, MBPP, StackEval, CodeContests) and industry data dumps to probe unseen robustness gaps.
- **Mind language balance.** Pair Java-heavy suites with datasets in JavaScript, C#, Go, Rust, and functional languages to test cross-language generalization.
- **Check for leakage.** Always publish hash lists or commit IDs so others can detect train–test overlap and avoid inflated robustness scores.
- **Report task strata.** Break down results by dataset and task; aggregated numbers can hide brittle model behavior on niche benchmarks.
- **Share data splits.** Release standard metamorphic splits (source vs. transformed) so future work can replicate and extend your analyses.

### 4.7 RQ7: Evaluation Metrics

To assess the effectiveness and robustness of deep code models under metamorphic testing, we systematically analyzed the evaluation metrics used in primary studies as summarized in Table 11, the most commonly used metric is **F1**, appearing in 12 of 45 studies (26.6%), with various forms including "F1 Score", and task-specific variants such as "ROB-F1" and "Gen-F1". This prevalence highlights the community's emphasis on balancing precision and recall tasks such as clone detection, vulnerability detection, and code completion.

**Accuracy** is the second most reported metric (28.89%), reflecting its simplicity and applicability to classification-based tasks. Metrics like **BLEU**, **ROUGE-L**, and **METEOR** are commonly used in generation tasks, especially code summarization and translation, revealing an influence of NLP style evaluation. Meanwhile, **ASR** (Attack Success Rate) and its variants are prevalent in adversarial robustness assessments, although inconsistently reported under multiple naming conventions.

We observe a high diversity in metric usage, with over 60 distinct metric labels, many of which appear in only one or two papers. This fragmentation indicates a lack of standardization and can limit the comparability between studies. Future work should aim to consolidate metric definitions and encourage the adoption of robust and interpretable measures for evaluating metamorphic transformations.

**Answer to RQ7.** The evaluation of metamorphic testing for deep code models relies on a diverse set of metrics. **F1** and **Accuracy** are the most common, each appearing in over a quarter of the studies, reflecting the community's emphasis on classification and correctness. Metrics such as **BLEU**, **ROUGE**, and **METEOR** are widely used for generation tasks, while **ASR** and its variants dominate adversarial robustness testing.

**Note to Researchers.** Despite steady progress, our survey reveals evaluation practices are still highly fragmented: synonymous metrics are inconsistently labeled, confidence intervals are rarely reported, and statistical significance is seldom tested.

Below are minimal, actionable recommendations that future studies can adopt to bring metric reporting onto a common, comparable footing:

- **Create a shared glossary.** Use a single label per metric (e.g., "F1", "ASR") and document the precise formula or implementation used.
- **Report uncertainty.** Include confidence intervals (e.g., via bootstrapping) for all metric values to reflect variance.
- **Add robustness-aware signals.** Report metrics alongside perturbation cost (e.g., number of tokens changed) and semantic similarity (e.g., CodeBLEU, BLEURT-Code).
- **Apply statistical tests.** Use significance testing (e.g., Wilcoxon, permutation test) before asserting superiority between approaches.
- **Release raw predictions.** Publish full model outputs so downstream researchers can replicate or extend metric analysis.

## 5 DISCUSSION AND FUTURE RESEARCH DIRECTIONS

A number of open research challenges emerge from this literature review, highlighting critical research gaps in metamorphic testing for deep code models. Below, we discuss the research challenges we have identified and possible future directions for researchers and practitioners.

### 5.1 Transformation Preferences and Intersections

*5.1.1 Intersections of Transformations with Metrics, Tasks, and Models.* Our analysis of transformation types in relation to evaluation metrics, tasks, and models reveals several consistent patterns. As shown in Figure 6, **Identifier Renaming**, **Dead Code Insertion**, and **Control Flow** transformations appear most frequently across different evaluation metrics. These transformations are syntactically safe, easy to apply, and widely compatible with static and dynamic analysis pipelines. Their popularity reflects both their implementation simplicity and their low risk of altering the semantic and behaviors of the programs being transformed.

A similar trend can be observed when transformations are mapped to downstream tasks. As depicted in Figure 7, Identifier Renaming is the most frequently used transformation in tasks such as clone detection, code summarization, and method name prediction. Likewise, Dead Code Insertion and Control Flow Transformations are broadly utilized across multiple tasks, suggesting their wide applicability and low disruption potential. Their effectiveness stems from the fact that they typically preserve both functionality and syntax, enabling seamless integration into diverse testing pipelines.

This trend continues across different families of models. As shown in Figure 8, these same transformation types are commonly used in evaluations involving both classical and modern architectures—including early models like LSTM and Code2Vec, as well as recent transformer-based models such as CodeBERT, GraphCodeBERT, CodeT5, and various GPT variants. The widespread adoption of these transformations across model types suggests their general compatibility and the existence of mature tool-chains for their automated application.

Table 11. Metrics used across primary studies, sorted by their fraction among all studies, along with the associated tasks and citing papers.

| Metric | Task(s) | Used by | Fraction (%) |
|---|---|---|---|
| F1 (all variants) | Method name prediction, Clone detection, Code completion, Vulnerability detection | [9, 65, 96, 90, 67, 71, 66, 60, 89, 79, 58, 69, 64, 92, 76, 78] | 35.56 |
| Accuracy | Vulnerability detection, Code summarization, Functionality classification, Clone detection | [39, 96, 90, 70, 67, 71, 94, 66, 91, 60, 59, 92, 73] | 28.89 |
| BLEU (all variants) | Code summarization, Code translation, Code completion | [8, 65, 77, 93, 71, 66, 91, 88, 69, 72, 87] | 24.44 |
| Attack Success Rate (ASR) (all forms) | Code completion, Code summarization, Vulnerability detection | [65, 95, 97, 74, 79, 62, 81, 85, 86, 83, 73, 57] | 26.67 |
| ROUGE / ROUGE-L | Code summarization, Comment generation | [77, 93, 71, 87] | 8.89 |
| METEOR | Code summarization, Comment generation | [77, 93, 71, 87] | 8.89 |
| Precision | Clone detection, Method name prediction | [90, 67, 94, 89, 64, 92] | 13.33 |
| Recall | Clone detection, Method name prediction | [90, 67, 94, 89, 64, 92] | 13.33 |
| MRR (all forms) | Code retrieval, Ranking tasks | [9, 93, 71, 91] | 8.89 |
| CodeBLEU | Code translation, Clone detection | [97, 72] | 4.44 |
| Exact Match (all forms) | Code translation | [71, 72] | 4.44 |
| Prediction Change Percentage | Functionality classification | [89, 58, 84] | 6.67 |
| Prediction Confidence Decrement (PCD) | Vulnerability detection | [63] | 2.22 |
| Delta-BLEU | Code summarization, Code translation | [66] | 2.22 |
| Gen-F1 | Clone detection (robustness) | [80] | 2.22 |
| ROB-F1 | Clone detection (robustness) | [80] | 2.22 |
| Pass@1 / Robust Pass@1 | Code generation | [72] | 2.22 |
| Mutation Effectiveness | General robustness testing | [70] | 2.22 |
| AUROC | Clone detection (contrastive learning) | [67] | 2.22 |
| Rate of Revealed Faults (RFR) | Defect prediction | [63] | 2.22 |

In contrast, transformation types such as **API changes**, **Function restructuring**, and **Comment or whitespace edits** remain significantly underutilized. These transformations are more complex to apply correctly, requiring contextual awareness to avoid altering program semantics or breaking syntactic constraints. Their lower adoption may also reflect a lack of robust, language-agnostic tools to support such transformations across different models and tasks.

Taken together, the three figures together indicate that researchers tend to use transformations that are easier to implement and syntactically conservative. While these transformations are valuable for initial robustness testing, relying mostly on them may lead to missing important weaknesses that only appear under more complex changes. Future work should seek to diversify transformation techniques and systematically apply underexplored types across a broader spectrum of models, languages, and tasks to uncover deeper model weaknesses.

| Transformation Type | F1 | ASR | Accuracy | BLEU | Precision | Recall | METEOR | ROUGE-L | MRR | CodeBLEU |
|---|---|---|---|---|---|---|---|---|---|---|
| Trivial | 8 | 4 | 4 | 3 | 2 | 2 | 0 | 0 | 1 | 0 |
| Identifier Renaming | 15 | 10 | 11 | 9 | 5 | 5 | 3 | 2 | 2 | 1 |
| Data | 7 | 3 | 6 | 4 | 2 | 2 | 1 | 1 | 0 | 1 |
| Control Flow | 9 | 7 | 8 | 5 | 3 | 3 | 1 | 1 | 2 | 1 |
| Function | 1 | 2 | 3 | 3 | 1 | 1 | 1 | 1 | 1 | 1 |
| Dead Code Insertion | 11 | 8 | 7 | 7 | 3 | 3 | 1 | 1 | 2 | 2 |
| API | 0 | 1 | 0 | 1 | 0 | 0 | 1 | 1 | 0 | 0 |
| Comment | 1 | 0 | 1 | 3 | 1 | 1 | 1 | 1 | 1 | 0 |
| Miscellaneous | 5 | 1 | 2 | 3 | 2 | 2 | 2 | 2 | 1 | 0 |

Fig. 6. Top Metrics and Transformation Types

| Transformation Type | Clone detection | Authorship attribution | Defect detection | Method name prediction | Functionality classification | Code summarization | Method Name Prediction | Code translation | Vulnerability prediction | Code classification |
|---|---|---|---|---|---|---|---|---|---|---|
| Trivial | 3 | 2 | 0 | 2 | 1 | 1 | 3 | 0 | 1 | 1 |
| Identifier Renaming | 9 | 5 | 4 | 5 | 5 | 4 | 4 | 2 | 2 | 2 |
| Data | 1 | 1 | 0 | 1 | 0 | 1 | 4 | 0 | 0 | 0 |
| Control Flow | 4 | 2 | 1 | 4 | 1 | 1 | 4 | 1 | 1 | 1 |
| Function | 0 | 1 | 0 | 0 | 0 | 0 | 0 | 0 | 0 | 0 |
| Dead Code Insertion | 6 | 3 | 3 | 4 | 2 | 2 | 2 | 1 | 0 | 1 |
| API | 1 | 2 | 0 | 0 | 1 | 0 | 0 | 0 | 1 | 0 |
| Comment | 2 | 0 | 1 | 0 | 0 | 0 | 0 | 1 | 0 | 0 |
| Miscellaneous | 1 | 1 | 0 | 0 | 0 | 0 | 1 | 0 | 0 | 0 |

Fig. 7. Top Tasks and Transformation Types

*5.1.2 Interpreting Transformation Preferences.* We observe a clear preference for transformation types that are simpler to implement and less likely to introduce semantic ambiguity. For example, **Identifier Renaming** appears in nearly 87% of studies, due to its ease of automation and minimal risk of breaking syntactic or functional correctness. Similarly, **Dead Code Insertion** and **Control Flow** transformations are frequently applied, as they can be implemented in a language-agnostic manner and typically preserve program output. These transformations are also more compatible with common static analysis tools and neural representations that are highly dependent on surface patterns or token structures.

The widespread adoption of these transformation types is further supported by their applicability to a broad range of models, such as **CodeBERT**, **GraphCodeBERT**, and tasks like **clone detection**, **summarization**, and **defect prediction**. These models often tokenize input code at the subword or identifier level, making them sensitive to perturbations in lexical elements while still being robust to changes that preserve the structure. This makes lightweight transformations particularly effective for probing model behavior without requiring deep instrumentation.

In contrast, transformation types such as **API-level changes**, **Function restructuring**, and **Comment or whitespace edits** are rarely adopted. This underuse may be attributed to the

| Transformation Types | CodeBERT | GraphCodeBERT | Code2Vec | Code2Seq | CodeT5 | LSTM | GGNN | ASTNN | LSTM-based Seq2Seq | Transformer-based Seq2Seq | Seq2Seq | TBCNN | ContraBERT | DeepTyper | GRU-based seq2seq | PLBART | GPT-3.5 | SequenceR | Recoder | CodeLlama | GPT-4 |
|---|---|---|---|---|---|---|---|---|---|---|---|---|---|---|---|---|---|---|---|---|---|
| Trivial | 4 | 2 | 7 | 5 | 2 | 2 | 1 | 2 | 1 | 0 | 0 | 0 | 0 | 1 | 1 | 1 | 1 | 1 | 1 | 0 | 0 |
| Identifier Renaming | 14 | 11 | 10 | 9 | 6 | 5 | 5 | 4 | 4 | 3 | 3 | 2 | 2 | 2 | 2 | 2 | 2 | 2 | 2 | 2 | 2 |
| Data | 2 | 3 | 5 | 7 | 2 | 2 | 2 | 0 | 1 | 1 | 2 | 0 | 1 | 2 | 0 | 1 | 0 | 2 | 2 | 1 | 1 |
| Control Flow | 7 | 6 | 7 | 6 | 4 | 2 | 4 | 2 | 1 | 0 | 2 | 0 | 2 | 1 | 1 | 2 | 2 | 1 | 1 | 2 | 2 |
| Function | 2 | 1 | 1 | 0 | 1 | 1 | 1 | 0 | 0 | 0 | 0 | 0 | 1 | 0 | 1 | 0 | 1 | 0 | 1 | 0 | 0 |
| Dead Code Insertion | 9 | 7 | 8 | 7 | 5 | 2 | 5 | 2 | 1 | 1 | 3 | 1 | 2 | 1 | 0 | 2 | 2 | 2 | 2 | 2 | 2 |
| API | 1 | 1 | 0 | 0 | 1 | 0 | 0 | 0 | 0 | 0 | 0 | 0 | 0 | 0 | 0 | 0 | 0 | 0 | 0 | 0 | 0 |
| Comment | 2 | 1 | 0 | 0 | 0 | 0 | 0 | 1 | 0 | 0 | 0 | 1 | 0 | 0 | 0 | 0 | 0 | 0 | 0 | 0 | 0 |
| Miscellaneous | 2 | 2 | 3 | 2 | 1 | 1 | 1 | 1 | 0 | 0 | 0 | 0 | 0 | 1 | 0 | 0 | 0 | 1 | 1 | 1 | 1 |

Victim Models

Fig. 8. Top Models and Transformation Types

complexity of ensuring semantic equivalence, particularly in cases where external dependencies or dynamic behaviors are involved. API modifications often require contextual understanding of call signatures and usage contracts, making them difficult to automate reliably. Changes in the function level can alter control dependencies or variable scopes, increasing the chance of unintended behavior. Regarding comment or whitespace mutations, their perceived lack of semantic impact might make them less attractive for testing deeper model robustness.

Overall, the current literature demonstrates a bias toward transformations that are both technically convenient and structurally conservative. Although these are valuable for initial robustness testing, their dominance risks narrowing the scope of evaluation and missing deeper semantic vulnerabilities. Future work could benefit from balancing these shallow transformations with more semantically rich or behavior-altering variations.

*5.1.3 Broadening Transformation Scope Beyond Semantic Preservation.* Our review focuses on semantic-preserving transformations, which align with the foundational goals of metamorphic testing (MT) to assess model robustness under input changes that should not affect expected behavior. This focus is also a direct reflection of the current state of research: the vast majority of primary studies in our review focuses on semantic-preserving transformations, particularly for robustness evaluation. However, we acknowledge that non-semantic-preserving transformations are critical in other code-related tasks such as translation, optimization, or synthesis, where output variation is not only acceptable but desired. For example, a transformation that changes an imperative code segment into a functional one may not preserve semantics line-by-line but could still yield correct behavior in the target domain.

Such transformations fall outside the traditional scope of MT but offer valuable new directions for extending evaluation frameworks, particularly for generative models that must operate across diverse specifications, styles, or languages. Our current taxonomy, grounded in prevailing literature, does not include these transformation types due to their limited adoption in existing MT-based studies. We therefore highlight this as a promising direction for future work, especially for extending evaluation strategies in tasks where semantic change is inherent.

## 5.2 Task and Model Coverage Gaps

*5.2.1 Underexplored Tasks.* Current research predominantly focuses on *binary classification* tasks such as clone detection (29%) and method name prediction (29%), which are closely aligned with NLP-based tasks. These tasks, given their established benchmarks and datasets, have been extensively evaluated in the context of metamorphic testing. However, other non-binary classification tasks such as code completion (9%) and code translation (6.5%), which are directly related to generating or modifying code remain underexplored. Similarly, tasks such as code repair (4.5%) and code search (6.5%) are also underrepresented despite their practical significance in real-world software engineering scenarios.

Hou et al. [53], in their systematic review of large language models for software engineering, emphasize that tasks such as code repair, code completion, code search, and especially code generation are among the most used and critical for practitioners. In particular, code generation is highlighted as a prioritized task in practice. However, within the domain of metamorphic testing for deep code models, these essential tasks are often neglected, creating a significant gap in research.

Researchers should address this imbalance, since deep code models should perform tasks like code repair and completion in a robust and reliable way to allow their use in real-world applications. Expanding robustness evaluations to encompass these underexplored yet essential tasks can significantly enhance the applicability and dependability of deep code models in modern software engineering practices. Additionally, Zhang et al. [120] showed that one of the primary tasks developers perform using tools like Copilot involves front-end development. Despite its practical importance, there is a clear lack of focus in the literature on evaluating the robustness of deep code models for front-end-related tasks.

*5.2.2 Model-Specific Evaluation Gaps.* Models such as CodeBERT [32] and GraphCodeBERT [103] are extensively tested against transformations, including identifier renaming and dead code insertion. However, more sophisticated transformations, such as API-level modifications or complex control flow alterations, are less explored. Furthermore, classic architectures, such as LSTM, remain vital, with studies like Zhang et al. [86] demonstrating their relevance under adversarial testing. Moreover, most studies did not include more recent models (e.g., CodeT5, Deepseek), which are reported to achieve better results in various tasks. Therefore, assessing the robustness of these new LLMs is an open research direction.

An equally critical gap is the lack of robustness evaluations for *closed-source models* frequently used in *semi-automated software engineering (SE) tasks*, such as *ChatGPT*, `Gemini`, and `GitHub Copilot`. These models differ from traditional deep code models, as they are *closed-source*, *proprietary*, and *chatbot-based*, integrating both code generation and *natural language understanding*. Therefore, metamorphic transformations could be applied not only to the code snippets themselves but also to the natural language component of prompts, introducing a new dimension in robustness testing. Investigating how these models handle such transformations is a promising research direction that could reveal new vulnerabilities and guide model providers in developing more resilient and robust AI-assisted programming tools.

While transformer-based models, like CodeT5, dominate current research, classic architectures, such as LSTM and Seq2Seq, remain integral to certain tasks. Continuing to evaluate and adapt these older architectures under modern robustness frameworks can provide valuable insights and ensure legacy systems benefit from advances in metamorphic testing methodologies.

The landscape of LLM4Code is expanding rapidly, with several high-impact models emerging from both industry and open-source communities. However, as shown in Table 12, many of these models are either absent or underrepresented in current academic research on metamorphic testing of deep code models. While models like Codex and GitHub Copilot have gained widespread adoption in software engineering practice, they appear in only a single primary study each. This highlights a growing divide between research focus and the tools actively used by practitioners. Similarly, models such as GPT-J, CodeParrot, CodeGen, and ChatGLM possess code generation or bilingual capabilities, yet remain largely unexplored in metamorphic robustness settings.

Table 12. Additional Notable Code Models: Rarely Studied or Recently Introduced

| Model | Institution | Notes |
|---|---|---|
| **Models Appearing in Only One Primary Study** | | |
| Codex [121] | OpenAI | Backbone of GitHub Copilot; trained on code. |
| GitHub Copilot [122] | GitHub / OpenAI | AI pair programmer; widely used in practice. |
| GPT-J [123] | EleutherAI | 6B parameter open LLM with code capability. |
| GPT-Neo [124] | EleutherAI | Open GPT-2-style model; used in low-resource code scenarios. |
| CodeParrot [125] | HuggingFace | Trained on Python from GitHub repos. |
| CodeGen [126] | Salesforce | Decoder-only model for multi-language code generation. |
| ChatGLM [127] | Tsinghua / Zhipu AI | Chinese bilingual LLM with code support. |
| StarCoder [128] | BigCode (HuggingFace + ServiceNow) | Open-source model trained on permissive GitHub code. |
| **Emerging Models Not Found in Our Primary Studies** | | |
| StarCoder2 [129] | BigCode + ServiceNow | New multilingual successor to StarCoder. |
| Codestral [130] | Mistral AI | Lightweight open model for structured code generation. |
| DeepSeek-Coder [131] | DeepSeek AI | 2T token pretraining; multilingual code support. |
| Magicoder [132] | Salesforce Research | Instruction-tuned version of CodeGen2. |
| CodeGemma [133] | Google DeepMind | Part of the Gemma family; open-source code specialization. |
| PolyCoder [134] | CMU | GPT-2 trained on C; interpretable and open. |
| Phi-2(Code-variant) [135] | Microsoft Research | Small model with strong reasoning and basic code capabilities. |

The second group in Table 12 includes emerging models not found in any of the primary studies included in our review. These include StarCoder2, Codestral, DeepSeek-Coder, Magicoder, CodeGemma, and Phi-2. Many of these models are not only recent but are specifically designed for code-related tasks, offering features such as multilingual support, instruction tuning, and lightweight deployment. Despite their potential, they have not yet been systematically evaluated for robustness using transformation-based testing techniques. Their absence from the literature presents both a gap and an opportunity. Evaluating these new models using rigorous metamorphic frameworks could significantly improve our understanding of their real-world reliability and guide future development.

### 5.3 Benchmarks and Language Diversity

*5.3.1 Language Diversity.* Java (33 studies), Python (19 studies), and C/C++ (24 studies combined) dominate the landscape of evaluated languages in the context of metamorphic testing for deep code models. Java's prominence can be attributed to its robust ecosystem and extensive use in software engineering tasks such as defect detection, clone detection, and code summarization. Python, given its versatility and frequent use in AI and data science, is another highly evaluated language, especially for tasks like code translation and summarization. However, despite their significance, languages such as JavaScript (5 studies), C# (4 studies), Go (4 studies), and PHP (3 studies) remain underexplored in metamorphic testing.

Interestingly, Zhang et al. [120] highlight that *Python and JavaScript* are the most frequently used languages in code generated by tools like GitHub Copilot, with *Node.js* being the most common technology among practitioners. Despite this, metamorphic robustness evaluations for JavaScript remain limited. Similarly, *C#*, widely used in enterprise applications through the *.NET framework*, and *Go*, known for its efficiency in *cloud native applications*, are rarely examined in robustness studies. Furthermore, languages like *Ruby* and *PHP*, which are fundamental to web development, receive little attention in metamorphic testing.

This mismatch between practitioner priorities and academic focus highlights the need for a broaden research efforts into less explored languages. Addressing these gaps would enhance the alignment between academic investigations and real-world software engineering practices. Expanding evaluations to include languages like *JavaScript, C#, Go,* and even *functional programming languages* (e.g., *Haskell*) could significantly improve the applicability of metamorphic testing research, ensuring that robustness assessments account for diverse programming paradigms.

*5.3.2 Dataset Limitations.* Datasets such as CodeSearchNet [112] and BigCloneBench [100] provide foundational benchmarks for many tasks. However, future work should prioritize creating benchmarks for languages and tasks that lack sufficient evaluation datasets, such as PHP, JavaScript, and front-end frameworks. Zhang et al. [120] highlight the use of a wide range of front end frameworks and technologies, including Vue, React, Next.js, Flutter, and Vanilla JS, in real world development with tools like Copilot. Future research should target benchmarks and use performance metrics that align with the practical priorities of developers. As shown in Table 13, several benchmarks with strong potential for metamorphic testing remain underutilized in the current literature. Datasets such as *VDISC*, *BFP*, and *VarMisuse* appeared in only a single primary study, despite offering well-structured scenarios for evaluating model behavior under transformations related to vulnerabilities, bug fixing, and semantic correctness. Additionally, widely recognized and practically relevant benchmarks such as *HumanEval*, *MBPP*, *StackEval*, and *CodeContests* have yet to be incorporated into metamorphic evaluation frameworks, even though their formats are well suited to prompt variation and semantic perturbation. The limited adoption of these datasets in metamorphic testing presents a missed opportunity to assess generalization, correctness, and resilience across diverse real world programming scenarios. Future work should broaden the use of these benchmarks to establish more comprehensive and transferable evaluation pipelines.

## 5.4 Beyond Robustness: Expanding Quality Objectives

*5.4.1 Opportunities for New Quality Objectives.* While our review shows that most existing work applies metamorphic testing (MT) to evaluate robustness, we believe there is a growing opportunity to extend MT toward other critical quality attributes. Based on the work of Yang et al. [47], quality objectives such as security, privacy, explainability, efficiency, and usability are increasingly relevant, especially as large language models for code are deployed in real-world development environments. MT provides a structured means to test model behavior under controlled, semantics-preserving changes and can be adapted to evaluate these broader dimensions. Below, we outline how MT can be designed to assess these additional quality attributes.

- **Security.** MT can simulate potential vulnerabilities by inserting no-op insecure patterns (e.g., insecure imports or unused cryptographic calls) and observing whether the model detects, flags, or reproduces these patterns in generated code. This is particularly useful for tasks like vulnerability detection or code completion.

Table 13. Additional Notable Benchmarks: Rarely Used or Recently Introduced

| Dataset | Origin | Task | MT Applicability |
| --- | --- | --- | --- |
| **Benchmarks Appearing in Only One Primary Study** | | | |
| VDISC [136] | Russell et al. | Vulnerability Detection | Allows testing model robustness by simulating security vulnerabilities. |
| BFP [137] | Tufano et al. | Program Repair | Enables robustness checks on automated bug-fix suggestions. |
| VarMisuse [31] | Allamanis et al. | Variable Misuse Detection | Useful for testing model sensitivity to semantic variable changes. |
| Defects4J [118] | Rjust et al. | Fault Localization | Supports fault simulation and validation of patch generalization. |
| LeetCode [138] | Wu et al. | General Code Generation | Enables prompt variation and paraphrasing for generation consistency tests. |
| NeuralCodeSum [104] | Ahmad et al. | Code Summarization | Applicable through syntactic/semantic transformations of input snippets. |
| **Emerging Benchmarks Not Found in Our Primary Studies** | | | |
| HumanEval [121] | OpenAI | Code Generation (NL-to-Code) | Facilitates testing robustness to prompt rewrites and code mutations. |
| MBPP [139] | Google | Python Programming Tasks | Supports controlled transformation of problem descriptions and solutions. |
| StackEval [140] | Prosus AI | Question-to-Code Generation | Designed for LLM evaluation with real-world prompts; MT can alter question phrasing and context. |
| CodeContests [141] | Salesforce | Competitive Programming | Enables testing model generalization under adversarial input variations. |
| Spider [142] | Yale NLP | NL-to-SQL Generation | Applicable by modifying NL queries or database schemas. |
| CodeNet [143] | IBM | Multi-language Code Generation | Supports cross-language robustness testing and perturbation-based analysis. |

- **Privacy.** MT can be used to evaluate whether minor changes to input data influence the likelihood of exposing sensitive tokens. For instance, adding neutral identifiers or unused strings may help detect whether the model leaks training data through memorization.
- **Explainability.** For interpretability-based tasks such as bug localization or code review suggestions, MT can introduce semantics-preserving changes and observe whether explanations, attention maps, or rationales remain consistent. Significant variation may indicate poor explainability.
- **Efficiency.** MT can test model performance under increasing input complexity by injecting deeply nested loops or long functions while maintaining semantics. This helps assess whether the model degrades gracefully in time or quality. Another approach is to enhance the efficiency of prompts and reduce inference time by applying metamorphic transformations that reduce the input size (i.e., the number of tokens), such as simplifying expressions or removing unnecessary comments or whitespace.
- **Usability.** Models should generate code that is understandable and useful to human developers. MT can apply human-friendly refactoring or adjust code style to test if the model maintains output quality and readability across variants.
- Table 14 summarizes how MT can be designed to assess each of these quality attributes in deep code models.

Table 14. Examples of Metamorphic Relations for Diverse Quality Objectives in Deep Code Models

| Quality Objective | Example Metamorphic Relation | Evaluation Goal |
|---|---|---|
| Robustness | Rename variables or reorder independent code blocks | Test consistency of model output across minor semantic-preserving changes |
| Security | Insert no-op vulnerable patterns or deprecated APIs | Assess whether the model flags or avoids generating insecure constructs |
| Privacy | Inject benign strings or rare identifiers into prompts | Evaluate if the model reveals memorized or sensitive training data |
| Explainability | Apply functionally equivalent changes (e.g., rename variables) | Check stability of rationale, explanation, or attention weights |
| Efficiency | Add deeply nested structures or increase input length | Examine degradation in output quality or latency under large inputs |
| Usability | Refactor code using best practices and readable styles | Assess whether the model maintains human oriented readability and formatting |

To consolidate these ideas, Table 14 presents illustrative metamorphic relations for each of the above quality objectives, showing how MT can be operationalized to assess a broader range of software quality concerns in deep code models.

## 5.5 Other Cross-Cutting Concerns

*5.5.1 Accuracy vs. Robustness.* With the recent advancements in developing better models, many researchers have introduced various AI-based techniques. To evaluate their techniques, they mostly focused on metrics related to the accuracy or performance of their proposed approaches without considering the robustness aspects of those models against adversarial attacks. We strongly recommend that researchers and model providers report both standard performance metrics and robustness evaluations, incorporating assessments such as stability under metamorphic transformations and susceptibility to perturbations. This will allow researchers to provide a more comprehensive evaluation of model performance beyond accuracy-related metrics.

*5.5.2 Data Leakage.* As noted by Sallou et al. [10], there is a strong connection between data leakage and metamorphic testing. Data leakage occurs when a model inadvertently learns patterns from unintended sources, such as pre-training data that overlaps with evaluation benchmarks. This can lead to overly optimistic performance estimates. For example, the recent study by Ramos et al. [144] provides evidence of data leakage related to software bugs benchmarks widely used in the related literature to assess the performance of automated tools for program repair, software testing, and fault localization.

To mitigate this issue, we further reinforce the recommendation by Sallou et al. [10], that is, to complement model performance evaluations with robustness assessments using metamorphic testing. The choice of metamorphic transformations should be task-specific to ensure meaningful robustness evaluations. For instance, identifier renaming or code obfuscation should not impact tasks like unit test generation or automated patching, as these tasks should rely on code structure rather than specific variable names. However, identifier names are crucial in NLP-related tasks such as method name prediction, and alternative metamorphic transformations should be considered to assess model sensitivity without introducing confounding factors.

*5.5.3 Explainability and Naturalness in Transformed Code.* Explainable AI (XAI) can play a critical role in visualizing and interpreting why specific transformations lead to particular changes in model outputs, making the testing process more transparent and actionable. However, the effectiveness of current XAI tools in this area remains unclear. There is a clear gap and an exciting opportunity to develop task-specific XAI methods for deep code models and rigorously test them. Moreover, evaluating the effectiveness of XAI methods in the context of metamorphic testing remains an open challenge, as robust metrics for measuring explainability in code models are still underdeveloped.

Using explainability metrics and tools to provide insights about new and meaningful transformations offers an exciting yet underexplored opportunity to understand and enhance model robustness. Additionally, explainability can improve the evaluation of naturalness in transformed code, ensuring that robustness improvements align with usability. In the context of regulations like the General Data Protection Regulation (GDPR) [145, 146], which emphasizes accountability, transparency, and user control over data in AI systems, explainability becomes even more critical for ensuring compliance and trustworthiness in deep code models. Researchers can better address robustness challenges while fostering trust and usability by integrating explainability into the metamorphic testing pipeline.

Moreover, [39, 88, 83, 86] mention that the naturalness of a transformed code snippet should also be taken into account. For this, [85] and [84] attempt to apply metamorphic testing with a secondary objective of making the transformed snippet appear more natural to humans. They both do a study with human participants to assess the improvement in the naturalness of code snippets generated by their method. Lastly, while white-box techniques [65, 82, 78] remain a focus, black-box approaches are gaining traction due to their practicality. For example, surrogate models [56, 81] and output-only feedback techniques [61, 63, 9] allow for broader applicability, making them well-suited for diverse scenarios.

*5.5.4 Industrial Adoption of MT.* While more AI-based technologies are increasingly being developed in industry, practitioners still face challenges with the quality assurance of their newly developed AI-based features. This is because, in many cases, the traditional test oracles are challenging to define. This is where metamorphic testing can play a crucial role in assessing the robustness of models in production environments. Despite its potential, the industrial adoption of metamorphic testing remains limited, with most research focusing on academic evaluations. Future studies should explore practical applications, integrating metamorphic testing into CI/CD pipelines, automated quality assurance frameworks, and AI model monitoring systems.

*5.5.5 Connections to Mutation Testing.* An underexplored yet promising opportunity lies in connecting metamorphic testing (MT) with mutation testing for traditional test code [147, 148] or AI models [149, 150], particularly through the notion of equivalent mutants [151]. Mutation testing introduces small code changes intended to alter behavior, helping assess the strength of test suites; equivalent mutants —those that do not affect observable behavior—pose a known challenge, as they are unintended and can skew test effectiveness metrics. In contrast, MT applies behavior-preserving transformations deliberately, based on expected properties of correct programs. While their purposes differ, MT transformations can resemble equivalent mutants in effect, since both leave observable behavior unchanged. Recognizing this connection points to opportunities for shared tooling, hybrid testing strategies, and more integrated evaluation pipelines that bridge traditional software testing with modern ML robustness analysis.

*5.5.6 Copilot Usage Patterns: Implications for Metamorphic Testing.* To better align metamorphic testing strategies with practical development workflows, we analyzed Copilot usage patterns using empirical data from a recent study [120]. Figure 9 summarizes the distribution of programming languages, IDEs, technologies, and function types commonly associated with Copilot. These insights inform the design of MT techniques for deep code models.

Languages such as JavaScript, Python, Java, and Kotlin dominate Copilot usage, indicating that MT research should prioritize support for these languages. As discussed in Section 5.3 (*Benchmarks*

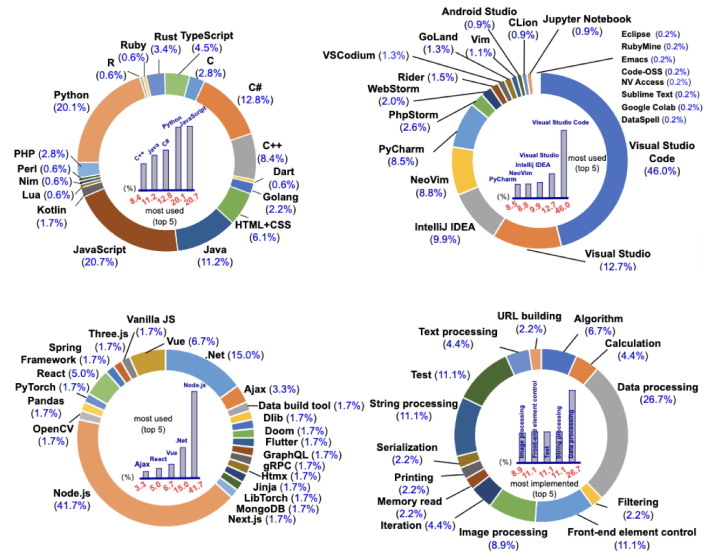

Fig. 9. Distribution of programming languages, IDEs, development technologies, and implemented functions used with GitHub Copilot, based on a empiricial study by Zhang et al. [120].

*and Language Diversity*), aligning test coverage with real-world language preferences enhances the external validity of MT-based evaluations.

On the tooling side, 26.7% of Copilot usage occurs within JetBrains IDEs—including IntelliJ IDEA, PyCharm, PhpStorm, and others—while Visual Studio Code accounts for 46%. These platforms represent strategic integration points for practical MT tooling. Embedding MT support directly into popular editors can facilitate adoption by developers and support real-time robustness analysis.

Technology ecosystems such as Node.js, .NET, and React are heavily used with Copilot, and generated functions frequently involve data processing, string manipulation, test code (11.1%), image handling, algorithm design, and UI control. These patterns suggest a need for domain-specific MT transformations, capable of preserving semantics across tasks like serialization, front-end behavior, or computation-heavy routines.

### 5.6 Road Map for Future Work

To guide future research in metamorphic testing of deep code models, we consolidate a structured roadmap that synthesizes recurring gaps and future opportunities identified across all RQs. This roadmap (Table 15) spans thirteen themes, each grounded in empirical findings from our review. It begins by urging the expansion of quality objectives beyond robustness to include security, privacy, fairness, and usability. It calls for diversification in transformation strategies, including non-semantic-preserving changes and mutation-MT synergy. Underexplored areas such as tasks (e.g., malware detection, memory prediction), model families (e.g., instruction-tuned and decoder-only LLMs), and programming languages (e.g., Rust, Kotlin, Solidity) are highlighted as priority domains. The roadmap also promotes innovation in benchmark design, standardization of evaluation metrics, and the integration of human factors, such as code readability and maintainability. Finally, it encourages explainability-guided testing, the automation of MT pipelines, the detection of overfitting, and the meta-learning of transferable transformations. Together, these directions form a comprehensive blueprint for advancing the reliability and generalizability of LLMs in software engineering tasks.

Table 15. Consolidated Road map for Metamorphic Testing of Deep Code Models

| Next Steps for Future Researchs |
| --- |
| **(1) Broaden Quality Objectives**<br>Move beyond robustness to cover *security, privacy, fairness, explainability, efficiency, and usability*. Inject insecure-but-compilable patterns, measure energy or latency under complexity-preserving loop edits, or test fairness via developer-introduced code variants. (RQ1, RQ2) |
| **(2) Diversify Transformation Space**<br>Introduce non-semantic-preserving transformations for generative tasks; treat MT variants as *equivalent mutants* to build synergy with mutation testing. Add API-level swaps, dependency edits, comment/whitespace tweaks, and restructure full functions. Publish semantic validators and AST tools. (RQ1) |
| **(3) Underexplored Code Tasks**<br>Expand MT to tasks such as code repair, completion, malware detection, memory region prediction, and test case generation. Develop task-specific metamorphic relations (e.g., API rewrites for translation). (RQ3) |
| **(4) Model Architecture Coverage**<br>Test instruction-tuned LLMs, including decoder-only models (e.g., StarCoder2, CodeLlama, Codestral, Magicoder, Phi-2), as well as black-box services like Copilot, ChatGPT, and Gemini. (RQ4) |
| **(5) Programming Language Diversity**<br>Expand to JavaScript, C#, Go, Rust, Kotlin, Swift, Solidity, SQL, etc. Release language-agnostic transformations and reusable parser plugins. Report per-language robustness results to reveal hidden gaps. (RQ5) |
| **(6) Benchmark Innovation**<br>Move beyond BigCloneBench and CodeSearchNet. Use newer corpora like HumanEval, MBPP, StackEval, Code-Contests, and VarMisuse. Release standard metamorphic splits and hash lists to detect leakage. (RQ6) |
| **(7) Metric Standardization**<br>Move beyond general metrics like accuracy or ASR and develop more task specific metrics for code based metamorphic testing. These should quantify aspects such as execution consistency, semantic preservation, and robustness. Always report confidence intervals, effect sizes, and release prediction logs with statistical tests for transparency and reproducibility. (RQ7) |
| **(8) Naturalness & Human Factors**<br>Quantify and evaluate how readable, maintainable, and acceptable code outputs are from a developer's perspective. Use human feedback to judge if perturbations remain valid and useful, and define thresholds for when transformations go too far. (RQ1, RQ7) |
| **(9) Explainability Guided MT**<br>Pair transformations with saliency maps or gradient attribution. Expose "why" the transformation fails and aid diagnosis beyond binary success/failure. (RQ2, Discussion) |
| **(10) Full Automation & CI/CD Integration**<br>Build one click MT pipelines (e.g., via GitHub Actions or Docker). Automate transformation generation, semantic checks, dashboards, and regression budgets. (All) |
| **(11) Overfitting & Leakage Checks**<br>Generate near duplicates of training data to detect memorization, license violations, or training bias. Pair with privacy audit tooling. (Discussion) |
| **(12) Cross-Task Generalisation**<br>Test if transformations crafted for one task (e.g., clone detection) transfer to another (e.g., vulnerability detection). Explore meta-learning of transferable MT strategies. (RQ3) |
| **(13) Mutation–MT Synergy**<br>Combine MT with mutation testing by labeling MT variants as equivalent mutants. Integrate both into shared evaluation pipelines with unified adequacy metrics. (Discussion) |

## 6 THREATS TO VALIDITY

*Limited Corpus Size.* While 45 papers may seem limited for a typical SLR, our focus on the niche and emerging area of metamorphic testing for deep code models justifies this scope. Besides, the domain of LLM4Code is still emerging, with the most significant developments occurring only in recent years. As a result, investigations into fairness, explainability, and other side properties remain underexplored. This temporal immaturity may constrain the availability of empirical evidence and limit the generalizability of observed trends.

*Construct and Classification Validity.* Metamorphic testing may be applied in practice without being explicitly labeled as such in the literature. This definitional ambiguity could result in undercoverage. To mitigate this threat, we explicitly included studies that used related terms such as "code transformation", "data augmentation", "robustness testing", and "semantic-preserving mutation", when they aligned with the conceptual criteria for metamorphic testing established in our study. Additionally, while we employed collaborative coding and cross-validation to ensure consistency in classification (e.g., task type, transformation strategy), some subjectivity may remain due to differences in interpretation or domain expertise.

*External Conclusions.* Our comparison with practitioners' perspectives, particularly from Zhang et al. [120], should be interpreted with caution, as their study was not specifically designed for testing deep code models. Additionally, while we primarily relied on evidence from our literature pool, we acknowledged newer models and benchmarks outside the original corpus in the discussion to highlight emerging trends.

## 7 CONCLUSION

This study presents a systematic literature review on metamorphic testing (MT) for deep code models, synthesizing insights from 45 peer-reviewed papers published between 2019 and 2024. We curated a comprehensive taxonomy of metamorphic transformation types and analyzed the evolution of application techniques, from early one-pass application techniques to more advanced optimization-based and feedback-driven methods. Beyond technique categorization, we systematically ranked the downstream tasks, programming languages, model architectures, data sets, and evaluation metrics that were the most frequently evaluated. This dual-layered analysis provides a structured overview of the current research landscape while highlighting critical, underexplored areas. Our findings revealed not only a growing methodological diversity but also recurring blind spots, such as limited task and language coverage, a lack of standard evaluation metrics, and insufficient attention to human factors. To address these, we proposed a research roadmap with actionable themes to advance metamorphic-based robustness evaluation of deep code models. We hope that this work serves as a foundation for future research and contributes toward more rigorous, generalizable, and practically applicable evaluation of code-focused large language models.

### ACKNOWLEDGMENTS

This work was conducted as part of the AI for Software Engineering (AI4SE) collaboration between JetBrains and Delft University of Technology. The authors gratefully acknowledge the financial support provided by JetBrains, which made this research possible.

### REFERENCES

[1] Hammond Pearce et al. "Examining zero-shot vulnerability repair with large language models". In: *2023 IEEE Symposium on Security and Privacy (SP)*. IEEE. 2023, pp. 2339–2356.

[2] Xin Zhou, Ting Zhang, and David Lo. "Large language model for vulnerability detection: Emerging results and future directions". In: *Proceedings of the 2024 ACM/IEEE 44th International Conference on Software Engineering: New Ideas and Emerging Results*. 2024, pp. 47–51.

[3] Yizheng Chen et al. "Diversevul: A new vulnerable source code dataset for deep learning based vulnerability detection". In: *Proceedings of the 26th International Symposium on Research in Attacks, Intrusions and Defenses*. 2023, pp. 654–668.

[4] Fang Liu et al. "Multi-task learning based pre-trained language model for code completion". In: *Proceedings of the 35th IEEE/ACM International Conference on Automated Software Engineering*. 2020, pp. 473–485.

[5] Daya Guo et al. "Longcoder: A long-range pre-trained language model for code completion". In: *International Conference on Machine Learning*. PMLR. 2023, pp. 12098–12107.

[6] Mohamad Khajezade et al. "Investigating the Efficacy of Large Language Models for Code Clone Detection". In: *Proceedings of the 32nd IEEE/ACM International Conference on Program Comprehension*. 2024, pp. 161–165.

[7] Tim Sonnekalb et al. "Generalizability of code clone detection on codebert". In: *Proceedings of the 37th IEEE/ACM International Conference on Automated Software Engineering*. 2022, pp. 1–3.

[8] Leonhard Applis, Annibale Panichella, and Arie van Deursen. "Assessing Robustness of ML-Based Program Analysis Tools using Metamorphic Program Transformations". In: *2021 36th IEEE/ACM International Conference on Automated Software Engineering (ASE)*. ISSN: 2643-1572. Nov. 2021, pp. 1377–1381. DOI: 10.1109/ASE51524.2021.9678706. URL: https://ieeexplore.ieee.org/abstract/document/9678706 (visited on 09/07/2024).

[9] Leonhard Applis, Annibale Panichella, and Ruben Marang. "Searching for Quality: Genetic Algorithms and Metamorphic Testing for Software Engineering ML". In: *Proceedings of the Genetic and Evolutionary Computation Conference*. GECCO '23. New York, NY, USA: Association for Computing Machinery, July 2023, pp. 1490–1498. ISBN: 9798400701191. DOI: 10.1145/3583131.3590379. URL: https://dl.acm.org/doi/10.1145/3583131.3590379 (visited on 09/10/2024).

[10] June Sallou, Thomas Durieux, and Annibale Panichella. "Breaking the silence: the threats of using llms in software engineering". In: *Proceedings of the 2024 ACM/IEEE 44th International Conference on Software Engineering: New Ideas and Emerging Results*. 2024, pp. 102–106.

[11] Jieke Shi et al. "Greening large language models of code". In: *Proceedings of the 46th International Conference on Software Engineering: Software Engineering in Society*. 2024, pp. 142–153.

[12] Annibale Panichella. "Metamorphic-Based Many-Objective Optimization of LLMs for Code-related Tasks". In: *Proceedings of the 47th International Conference on Software Engineering*. 2025.

[13] Hammond Pearce et al. "Asleep at the keyboard? assessing the security of github copilot's code contributions". In: *2022 IEEE Symposium on Security and Privacy (SP)*. IEEE. 2022, pp. 754–768.

[14] Badhan Chandra Das, M Hadi Amini, and Yanzhao Wu. "Security and privacy challenges of large language models: A survey". In: *ACM Computing Surveys* (2024).

[15] Zhou Yang et al. "Natural attack for pre-trained models of code". In: *Proceedings of the 44th International Conference on Software Engineering*. 2022, pp. 1482–1493.

[16] Valentin JM Manès et al. "The art, science, and engineering of fuzzing: A survey". In: *IEEE Transactions on Software Engineering* 47.11 (2019), pp. 2312–2331.

[17] Phil McMinn. "Search-based software test data generation: a survey". In: *Software testing, Verification and reliability* 14.2 (2004), pp. 105–156.

[18] Earl T. Barr et al. "The Oracle Problem in Software Testing: A Survey". In: *IEEE Transactions on Software Engineering* 41.5 (May 2015). Conference Name: IEEE Transactions on Software Engineering, pp. 507–525. ISSN: 1939-3520. DOI: 10.1109/TSE.2014.2372785. URL: https://ieeexplore.ieee.org/document/6963470 (visited on 09/19/2024).

[19] Sergio Segura et al. "A Survey on Metamorphic Testing". In: *IEEE Transactions on Software Engineering* 42.9 (Sept. 2016). Conference Name: IEEE Transactions on Software Engineering, pp. 805–824. ISSN: 1939-3520. DOI: 10.1109/TSE.2016.2532875. URL: https://ieeexplore.ieee.org/abstract/document/7422146 (visited on 09/08/2024).

[20] WK Chan, Shing Chi Cheung, and Karl RPH Leung. "Towards a metamorphic testing methodology for service-oriented software applications". In: *Fifth International Conference on Quality Software (QSIC'05)*. IEEE. 2005, pp. 470–476.

[21] Chang-Ai Sun et al. "Metamorphic Testing of Image Processing Applications: A General Framework and Optimization Strategies". In: *Proceedings of the 9th ACM International Workshop on Metamorphic Testing*. 2024, pp. 26–33.

[22] Qiuming Tao et al. "An automatic testing approach for compiler based on metamorphic testing technique". In: *2010 Asia Pacific Software Engineering Conference*. IEEE. 2010, pp. 270–279.

[23] Tsong Yueh Chen et al. "Metamorphic testing: A review of challenges and opportunities". In: *ACM Computing Surveys (CSUR)* 51.1 (2018), pp. 1–27.

[24] Sergio Segura and Zhi Quan Zhou. "Metamorphic testing 20 years later: A hands-on introduction". In: *Proceedings of the 40th International Conference on Software Engineering: Companion Proceeedings*. 2018, pp. 538–539.

[25] Nazanin Bayati Chaleshtari et al. "Metamorphic testing for web system security". In: *IEEE Transactions on Software Engineering* 49.6 (2023), pp. 3430–3471.

[26] Faqeer Ur Rehman and Madhusudan Srinivasan. "Metamorphic Testing For Machine Learning: Applicability, Challenges, and Research Opportunities". In: *2023 IEEE International Conference On Artificial Intelligence Testing (AITest)*. IEEE. 2023, pp. 34–39.

[27] Morteza Pourreza Shahri et al. "Metamorphic testing for quality assurance of protein function prediction tools". In: *2019 IEEE International Conference On Artificial Intelligence Testing (AITest)*. IEEE. 2019, pp. 140–148.

[28] Diogo Moreira, Ana Paula Furtado, and Sidney Nogueira. "Testing acoustic scene classifiers using Metamorphic Relations". In: *2020 IEEE International Conference On Artificial Intelligence Testing (AITest)*. IEEE. 2020, pp. 47–54.

[29] Uri Alon et al. "code2vec: learning distributed representations of code". In: *Implementation, data and a trained model for the code2vec paper* 3.POPL (Jan. 2019), 40:1–40:29. DOI: 10.1145/3290353. URL: https://dl.acm.org/doi/10.1145/3290353 (visited on 10/30/2024).

[30] Uri Alon et al. "code2seq: Generating Sequences from Structured Representations of Code". In: *International Conference on Learning Representations*. 2019. URL: https://openreview.net/forum?id=H1gKYo09tX.

[31] Miltiadis Allamanis, Marc Brockschmidt, and Mahmoud Khademi. *Learning to Represent Programs with Graphs*. arXiv:1711.00740. May 2018. DOI: 10.48550/arXiv.1711.00740. URL: http://arxiv.org/abs/1711.00740 (visited on 10/30/2024).

[32] Zhangyin Feng et al. *CodeBERT: A Pre-Trained Model for Programming and Natural Languages*. arXiv:2002.08155. Sept. 2020. DOI: 10.48550/arXiv.2002.08155. URL: http://arxiv.org/abs/2002.08155 (visited on 10/30/2024).

[33] Yue Wang et al. *CodeT5: Identifier-aware Unified Pre-trained Encoder-Decoder Models for Code Understanding and Generation.* arXiv:2109.00859. Sept. 2021. DOI: 10.48550/arXiv.2109.00859. URL: http://arxiv.org/abs/2109.00859 (visited on 10/30/2024).

[34] Vincent J. Hellendoorn et al. "Deep learning type inference". In: *Proceedings of the 2018 26th ACM Joint Meeting on European Software Engineering Conference and Symposium on the Foundations of Software Engineering.* ESEC/FSE 2018. New York, NY, USA: Association for Computing Machinery, Oct. 2018, pp. 152–162. ISBN: 978-1-4503-5573-5. DOI: 10.1145/3236024.3236051. URL: https://dl.acm.org/doi/10.1145/3236024.3236051 (visited on 10/30/2024).

[35] Josh Achiam et al. "Gpt-4 technical report". In: *arXiv preprint arXiv:2303.08774* (2023).

[36] Aram Bahrini et al. "ChatGPT: Applications, Opportunities, and Threats". In: *2023 Systems and Information Engineering Design Symposium (SIEDS).* Apr. 2023, pp. 274–279. DOI: 10.1109/SIEDS58326.2023.10137850. URL: https://ieeexplore.ieee.org/abstract/document/10137850 (visited on 10/30/2024).

[37] Hugo Touvron et al. "Llama: Open and efficient foundation language models". In: *arXiv preprint arXiv:2302.13971* (2023).

[38] T. Y. Chen, S. C. Cheung, and S. M. Yiu. *Metamorphic Testing: A New Approach for Generating Next Test Cases.* arXiv:2002.12543 [cs]. Feb. 2020. DOI: 10.48550/arXiv.2002.12543. URL: http://arxiv.org/abs/2002.12543 (visited on 09/18/2024).

[39] Pavol Bielik and Martin Vechev. "Adversarial Robustness for Code". en. In: *Proceedings of the 37th International Conference on Machine Learning.* ISSN: 2640-3498. PMLR, Nov. 2020, pp. 896–907. URL: https://proceedings.mlr.press/v119/bielik20a.html (visited on 09/12/2024).

[40] Pengyu Xue et al. "Exploring and Lifting the Robustness of LLM-powered Automated Program Repair with Metamorphic Testing". In: *arXiv preprint arXiv:2410.07516* (2024).

[41] Ziyu Li et al. "Evaluating LLM's Code Reading Abilities in Big Data Contexts using Metamorphic Testing". In: *2023 9th International Conference on Big Data and Information Analytics (BigDIA).* 2023, pp. 232–239. DOI: 10.1109/BigDIA60676.2023.10429345.

[42] Harishwar Reddy, Madhusudan Srinivasan, and Upulee Kanewala. "Metamorphic Testing for Fairness Evaluation in Large Language Models: Identifying Intersectional Bias in LLaMA and GPT". In: *arXiv preprint arXiv:2504.07982* (2025).

[43] Guna Sekaran Jaganathan, Indika Kahanda, and Upulee Kanewala. "Metamorphic Testing for robustness and fairness evaluation of LLM-based automated ICD coding applications". In: *Smart Health* (2025), p. 100564.

[44] Sangwon Hyun, Mingyu Guo, and M Ali Babar. "METAL: Metamorphic Testing Framework for Analyzing Large-Language Model Qualities". In: *2024 IEEE Conference on Software Testing, Verification and Validation (ICST).* IEEE. 2024, pp. 117–128.

[45] Muxin Pu et al. "Fairness evaluation in deepfake detection models using metamorphic testing". In: *Proceedings of the 7th international workshop on metamorphic testing.* 2022, pp. 7–14.

[46] Zhahao Li et al. "Detecting Bias in LLMs' Natural Language Inference Using Metamorphic Testing". In: *2024 IEEE 24th International Conference on Software Quality, Reliability, and Security Companion (QRS-C).* IEEE. 2024, pp. 31–37.

[47] Zhou Yang et al. "Robustness, security, privacy, explainability, efficiency, and usability of large language models for code". In: *arXiv preprint arXiv:2403.07506* (2024).

[48] Huangzhao Zhang et al. "Towards robustness of deep program processing models—detection, estimation, and enhancement". In: *ACM Transactions on Software Engineering and Methodology (TOSEM)* 31.3 (2022), pp. 1–40.

[49] Zohreh Aghababaeyan et al. "Black-box testing of deep neural networks through test case diversity". In: *IEEE Transactions on Software Engineering* 49.5 (2023), pp. 3182–3204.

[50] Zhen Li et al. "A comparative study of adversarial training methods for neural models of source code". In: *Future Generation Computer Systems* 142 (May 2023), pp. 165–181. ISSN: 0167-739X. DOI: 10.1016/j.future.2022.12.030. URL: https://www.sciencedirect.com/science/article/pii/S0167739X22004332 (visited on 09/16/2024).

[51] Yubin Qu, Song Huang, and Yongming Yao. "A survey on robustness attacks for deep code models". en. In: *Automated Software Engineering* 31.2 (Aug. 2024), p. 65. ISSN: 1573-7535. DOI: 10.1007/s10515-024-00464-7. URL: https://doi.org/10.1007/s10515-024-00464-7 (visited on 09/11/2024).

[52] Md Rafiqul Islam Rabin. "Methodologies for Evaluating and Interpreting Neural Code Intelligence Models". eng. In: (Apr. 2023). URL: https://hdl.handle.net/10657/15039 (visited on 09/18/2024).

[53] Xinyi Hou et al. "Large language models for software engineering: A systematic literature review". In: *ACM Transactions on Software Engineering and Methodology* 33.8 (2024), pp. 1–79.

[54] Jane Webster and Richard T. Watson. "Analyzing the Past to Prepare for the Future: Writing a Literature Review". In: *MIS Quarterly* 26.2 (2002). Publisher: Management Information Systems Research Center, University of Minnesota, pp. xiii–xxiii. ISSN: 0276-7783. URL: https://www.jstor.org/stable/4132319 (visited on 09/08/2024).

[55] Barbara Kitchenham et al. "Systematic literature reviews in software engineering–a systematic literature review". In: *Information and software technology* 51.1 (2009), pp. 7–15.

[56] Qianjun Liu et al. "A Practical Black-Box Attack on Source Code Authorship Identification Classifiers". In: *IEEE Transactions on Information Forensics and Security* 16 (2021). Conference Name: IEEE Transactions on Information Forensics and Security, pp. 3620–3633. ISSN: 1556-6021. DOI: 10.1109/TIFS.2021.3080507. URL: https://ieeexplore.ieee.org/document/9454564 (visited on 10/03/2024).

[57] Erwin Quiring, Alwin Maier, and Konrad Rieck. *Misleading Authorship Attribution of Source Code using Adversarial Learning.* arXiv:1905.12386 [cs, stat]. May 2019. DOI: 10.48550/arXiv.1905.12386. URL: http://arxiv.org/abs/1905.12386 (visited on 09/13/2024).

[58] Jacob M. Springer, Bryn Marie Reinstadler, and Una-May O'Reilly. *STRATA: Simple, Gradient-Free Attacks for Models of Code.* arXiv:2009.13562 [cs, stat]. Aug. 2021. DOI: 10.48550/arXiv.2009.13562. URL: http://arxiv.org/abs/2009.13562 (visited on 09/18/2024).

[59] Md Rafiqul Islam Rabin, Ke Wang, and Mohammad Amin Alipour. *Testing Neural Program Analyzers.* arXiv:1908.10711 [cs, stat]. Sept. 2019. DOI: 10.48550/arXiv.1908.10711. URL: http://arxiv.org/abs/1908.10711 (visited on 09/12/2024).

[60] Md Rafiqul Islam Rabin and Mohammad Amin Alipour. *Evaluation of Generalizability of Neural Program Analyzers under Semantic-Preserving Transformations.* arXiv:2004.07313 [cs]. Mar. 2021. DOI: 10.48550/arXiv.2004.07313. URL: http://arxiv.org/abs/2004.07313 (visited on 09/13/2024).

[61] Fengjuan Gao, Yu Wang, and Ke Wang. "Discrete Adversarial Attack to Models of Code". In: *Proc. ACM Program. Lang.* 7.PLDI (June 2023), 113:172–113:195. DOI: 10.1145/3591227. URL: https://dl.acm.org/doi/10.1145/3591227 (visited on 09/10/2024).

[62] Junfeng Tian et al. "Generating Adversarial Examples of Source Code Classification Models via Q-Learning-Based Markov Decision Process". In: *2021 IEEE 21st International Conference on Software Quality, Reliability and Security (QRS).* ISSN: 2693-9177. Dec. 2021, pp. 807–818. DOI: 10.1109/QRS54544.2021.00090. URL: https://ieeexplore.ieee.org/document/9724884 (visited on 09/18/2024).

[63] Zhao Tian, Junjie Chen, and Zhi Jin. "Code Difference Guided Adversarial Example Generation for Deep Code Models". In: *2023 38th IEEE/ACM International Conference on Automated Software Engineering (ASE)*. ISSN: 2643-1572. Sept. 2023, pp. 850–862. DOI: 10.1109/ASE56229.2023.00149. URL: https://ieeexplore.ieee.org/document/10298520 (visited on 09/11/2024).

[64] Weiwei Zhang et al. "Challenging Machine Learning-Based Clone Detectors via Semantic-Preserving Code Transformations". In: *IEEE Transactions on Software Engineering* 49.5 (May 2023). Conference Name: IEEE Transactions on Software Engineering, pp. 3052–3070. ISSN: 1939-3520. DOI: 10.1109/TSE.2023.3240118. URL: https://ieeexplore.ieee.org/document/10028657 (visited on 09/12/2024).

[65] Penglong Chen et al. "Generating Adversarial Source Programs Using Important Tokens-based Structural Transformations". In: *2022 26th International Conference on Engineering of Complex Computer Systems (ICECCS)*. Mar. 2022, pp. 173–182. DOI: 10.1109/ICECCS54210.2022.00029. URL: https://ieeexplore.ieee.org/document/9763729 (visited on 09/18/2024).

[66] Dexin Liu and Shikun Zhang. "ALANCA: Active Learning Guided Adversarial Attacks for Code Comprehension on Diverse Pre-trained and Large Language Models". In: *2024 IEEE International Conference on Software Analysis, Evolution and Reengineering (SANER)*. ISSN: 2640-7574. Mar. 2024, pp. 602–613. DOI: 10.1109/SANER60148.2024.00067. URL: https://ieeexplore.ieee.org/abstract/document/10589851 (visited on 09/18/2024).

[67] Paras Jain et al. "Contrastive Code Representation Learning". In: *Proceedings of the 2021 Conference on Empirical Methods in Natural Language Processing*. arXiv:2007.04973 [cs, stat]. 2021, pp. 5954–5971. DOI: 10.18653/v1/2021.emnlp-main.482. URL: http://arxiv.org/abs/2007.04973 (visited on 09/16/2024).

[68] Hongliang Ge et al. "RobustNPR: Evaluating the robustness of neural program repair models". en. In: *Journal of Software: Evolution and Process* 36.4 (2024). _eprint: https://onlinelibrary.wiley.com/doi/pdf/10.1002/smr.2586, e2586. ISSN: 2047-7481. DOI: 10.1002/smr.2586. URL: https://onlinelibrary.wiley.com/doi/abs/10.1002/smr.2586 (visited on 09/18/2024).

[69] Moshi Wei et al. "CoCoFuzzing: Testing Neural Code Models With Coverage-Guided Fuzzing". In: *IEEE Transactions on Reliability* 72.3 (Sept. 2023). Conference Name: IEEE Transactions on Reliability, pp. 1276–1289. ISSN: 1558-1721. DOI: 10.1109/TR.2022.3208239. URL: https://ieeexplore.ieee.org/abstract/document/9916170 (visited on 09/18/2024).

[70] Ashish Hooda et al. *Do Large Code Models Understand Programming Concepts? A Black-box Approach*. arXiv:2402.05980 [cs]. Feb. 2024. DOI: 10.48550/arXiv.2402.05980. URL: http://arxiv.org/abs/2402.05980 (visited on 09/11/2024).

[71] Yiyang Li, Hongqiu Wu, and Hai Zhao. *Semantic-Preserving Adversarial Code Comprehension*. arXiv:2209.05130 [cs]. Sept. 2022. DOI: 10.48550/arXiv.2209.05130. URL: http://arxiv.org/abs/2209.05130 (visited on 09/16/2024).

[72] Guang Yang et al. *Assessing and Improving Syntactic Adversarial Robustness of Pre-trained Models for Code Translation*. arXiv:2310.18587 [cs]. Oct. 2023. DOI: 10.48550/arXiv.2310.18587. URL: http://arxiv.org/abs/2310.18587 (visited on 09/16/2024).

[73] Chi Zhang et al. *Transfer Attacks and Defenses for Large Language Models on Coding Tasks*. arXiv:2311.13445 [cs]. Nov. 2023. DOI: 10.48550/arXiv.2311.13445. URL: http://arxiv.org/abs/2311.13445 (visited on 09/16/2024).

[74] Thanh-Dat Nguyen et al. *Adversarial Attacks on Code Models with Discriminative Graph Patterns*. arXiv:2308.11161 [cs]. Aug. 2023. DOI: 10.48550/arXiv.2308.11161. URL: http://arxiv.org/abs/2308.11161 (visited on 09/16/2024).

[75] Thanh Le-Cong et al. *Evaluating Program Repair with Semantic-Preserving Transformations: A Naturalness Assessment*. arXiv:2402.11892 [cs]. Feb. 2024. DOI: 10.48550/arXiv.2402.11892. URL: http://arxiv.org/abs/2402.11892 (visited on 09/17/2024).

[76]  Kexin Pei et al. *Exploiting Code Symmetries for Learning Program Semantics*. arXiv:2308.03312 [cs]. Sept. 2024. DOI: 10.48550/arXiv.2308.03312. URL: http://arxiv.org/abs/2308.03312 (visited on 09/17/2024).

[77]  Xi Ding et al. "Adversarial Attack and Robustness Improvement on Code Summarization". In: *Proceedings of the 28th International Conference on Evaluation and Assessment in Software Engineering*. EASE '24. New York, NY, USA: Association for Computing Machinery, June 2024, pp. 17–27. ISBN: 9798400717017. DOI: 10.1145/3661167.3661173. URL: https://dl.acm.org/doi/10.1145/3661167.3661173 (visited on 09/18/2024).

[78]  Jordan Henkel et al. "Semantic Robustness of Models of Source Code". In: *2022 IEEE International Conference on Software Analysis, Evolution and Reengineering (SANER)*. ISSN: 1534-5351. Mar. 2022, pp. 526–537. DOI: 10.1109/SANER53432.2022.00070. URL: https://ieeexplore.ieee.org/document/9825895 (visited on 09/12/2024).

[79]  Shashank Srikant et al. *Generating Adversarial Computer Programs using Optimized Obfuscations*. arXiv:2103.11882 [cs]. Mar. 2021. DOI: 10.48550/arXiv.2103.11882. URL: http://arxiv.org/abs/2103.11882 (visited on 09/11/2024).

[80]  Jinghan Jia et al. "ClawSAT: Towards Both Robust and Accurate Code Models". In: *2023 IEEE International Conference on Software Analysis, Evolution and Reengineering (SANER)*. ISSN: 2640-7574. Mar. 2023, pp. 212–223. DOI: 10.1109/SANER56733.2023.00029. URL: https://ieeexplore.ieee.org/document/10123554 (visited on 09/11/2024).

[81]  Yulong Yang et al. "Exploiting the Adversarial Example Vulnerability of Transfer Learning of Source Code". In: *IEEE Transactions on Information Forensics and Security* 19 (2024). Conference Name: IEEE Transactions on Information Forensics and Security, pp. 5880–5894. ISSN: 1556-6021. DOI: 10.1109/TIFS.2024.3402153. URL: https://ieeexplore.ieee.org/document/10531252 (visited on 09/18/2024).

[82]  Noam Yefet, Uri Alon, and Eran Yahav. "Adversarial examples for models of code". In: *Proc. ACM Program. Lang.* 4.OOPSLA (Nov. 2020), 162:1–162:30. DOI: 10.1145/3428230. URL: https://dl.acm.org/doi/10.1145/3428230 (visited on 09/10/2024).

[83]  Huangzhao Zhang et al. "Generating Adversarial Examples for Holding Robustness of Source Code Processing Models". en. In: *Proceedings of the AAAI Conference on Artificial Intelligence* 34.01 (Apr. 2020). Number: 01, pp. 1169–1176. ISSN: 2374-3468. DOI: 10.1609/aaai.v34i01.5469. URL: https://ojs.aaai.org/index.php/AAAI/article/view/5469 (visited on 09/12/2024).

[84]  Shasha Zhou et al. "Evolutionary Multi-objective Optimization for Contextual Adversarial Example Generation". In: *Proc. ACM Softw. Eng.* 1.FSE (July 2024), 101:2285–101:2308. DOI: 10.1145/3660808. URL: https://dl.acm.org/doi/10.1145/3660808 (visited on 09/10/2024).

[85]  Zhou Yang et al. "Natural attack for pre-trained models of code". In: *Proceedings of the 44th International Conference on Software Engineering*. ICSE '22. New York, NY, USA: Association for Computing Machinery, July 2022, pp. 1482–1493. ISBN: 978-1-4503-9221-1. DOI: 10.1145/3510003.3510146. URL: https://doi.org/10.1145/3510003.3510146 (visited on 09/10/2024).

[86]  Huangzhao Zhang et al. "CodeBERT-Attack: Adversarial attack against source code deep learning models via pre-trained model". en. In: *Journal of Software: Evolution and Process* 36.3 (2024). _eprint: https://onlinelibrary.wiley.com/doi/pdf/10.1002/smr.2571, e2571. ISSN: 2047-7481. DOI: 10.1002/smr.2571. URL: https://onlinelibrary.wiley.com/doi/abs/10.1002/smr.2571 (visited on 09/18/2024).

[87]  Yu Zhou et al. "Adversarial Robustness of Deep Code Comment Generation". In: *ACM Trans. Softw. Eng. Methodol.* 31.4 (July 2022), 60:1–60:30. ISSN: 1049-331X. DOI: 10.1145/3501256. URL: https://doi.org/10.1145/3501256 (visited on 09/10/2024).

[88] Zongjie Li et al. *CCTEST: Testing and Repairing Code Completion Systems.* arXiv:2208.08289 [cs]. May 2023. DOI: 10.48550/arXiv.2208.08289. URL: http://arxiv.org/abs/2208.08289 (visited on 09/16/2024).

[89] Md Rafiqul Islam Rabin et al. "On the generalizability of Neural Program Models with respect to semantic-preserving program transformations". In: *Information and Software Technology* 135 (July 2021), p. 106552. ISSN: 0950-5849. DOI: 10.1016/j.infsof.2021.106552. URL: https://www.sciencedirect.com/science/article/pii/S0950584921000379 (visited on 09/12/2024).

[90] Shuzheng Gao et al. "Two Sides of the Same Coin: Exploiting the Impact of Identifiers in Neural Code Comprehension". In: *Proceedings of the 45th International Conference on Software Engineering.* ICSE '23. Melbourne, Victoria, Australia: IEEE Press, July 2023, pp. 1933–1945. ISBN: 978-1-66545-701-9. DOI: 10.1109/ICSE48619.2023.00164. URL: https://doi.org/10.1109/ICSE48619.2023.00164 (visited on 09/10/2024).

[91] Shangqing Liu et al. "ContraBERT: Enhancing Code Pre-Trained Models via Contrastive Learning". In: *Proceedings of the 45th International Conference on Software Engineering.* ICSE '23. Melbourne, Victoria, Australia: IEEE Press, July 2023, pp. 2476–2487. ISBN: 978-1-66545-701-9. DOI: 10.1109/ICSE48619.2023.00207. URL: https://doi.org/10.1109/ICSE48619.2023.00207 (visited on 09/10/2024).

[92] Huangzhao Zhang et al. "Towards Robustness of Deep Program Processing Models—Detection, Estimation, and Enhancement". In: *ACM Trans. Softw. Eng. Methodol.* 31.3 (Apr. 2022), 50:1–50:40. ISSN: 1049-331X. DOI: 10.1145/3511887. URL: https://dl.acm.org/doi/10.1145/3511887 (visited on 09/12/2024).

[93] Yaoxian Li et al. *A Closer Look into Transformer-Based Code Intelligence Through Code Transformation: Challenges and Opportunities.* arXiv:2207.04285 [cs]. July 2022. DOI: 10.48550/arXiv.2207.04285. URL: http://arxiv.org/abs/2207.04285 (visited on 09/16/2024).

[94] Zhen Li et al. *Towards Making Deep Learning-based Vulnerability Detectors Robust.* arXiv:2108.00669 [cs]. Aug. 2021. DOI: 10.48550/arXiv.2108.00669. URL: http://arxiv.org/abs/2108.00669 (visited on 09/16/2024).

[95] Heng Li et al. *Black-box Adversarial Example Attack towards FCG Based Android Malware Detection under Incomplete Feature Information.* arXiv:2303.08509 [cs]. Mar. 2023. DOI: 10.48550/arXiv.2303.08509. URL: http://arxiv.org/abs/2303.08509 (visited on 09/18/2024).

[96] Claudio Ferretti and Martina Saletta. "Deceiving neural source code classifiers: finding adversarial examples with grammatical evolution". In: *Proceedings of the Genetic and Evolutionary Computation Conference Companion.* GECCO '21. New York, NY, USA: Association for Computing Machinery, July 2021, pp. 1889–1897. ISBN: 978-1-4503-8351-6. DOI: 10.1145/3449726.3463222. URL: https://dl.acm.org/doi/10.1145/3449726.3463222 (visited on 09/18/2024).

[97] CheolWon Na, YunSeok Choi, and Jee-Hyong Lee. "DIP: Dead code Insertion based Blackbox Attack for Programming Language Model". In: *Proceedings of the 61st Annual Meeting of the Association for Computational Linguistics (Volume 1: Long Papers).* Ed. by Anna Rogers, Jordan Boyd-Graber, and Naoaki Okazaki. Toronto, Canada: Association for Computational Linguistics, July 2023, pp. 7777–7791. DOI: 10.18653/v1/2023.acl-long.430. URL: https://aclanthology.org/2023.acl-long.430 (visited on 09/13/2024).

[98] Kalyanmoy Deb et al. "A fast elitist non-dominated sorting genetic algorithm for multi-objective optimization: NSGA-II". In: *Parallel Problem Solving from Nature PPSN VI: 6th International Conference Paris, France, September 18–20, 2000 Proceedings 6.* Springer. 2000, pp. 849–858.

[99]    Maciej Świechowski et al. "Monte Carlo tree search: A review of recent modifications and applications". In: *Artificial Intelligence Review* 56.3 (2023), pp. 2497–2562.

[100]   Jeffrey Svajlenko and Chanchal K. Roy. "Evaluating clone detection tools with BigCloneBench". In: *2015 IEEE International Conference on Software Maintenance and Evolution (ICSME)*. Sept. 2015, pp. 131–140. DOI: 10.1109/ICSM.2015.7332459. URL: https://ieeexplore.ieee.org/abstract/document/7332459 (visited on 10/30/2024).

[101]   Lili Mou et al. "Convolutional Neural Networks over Tree Structures for Programming Language Processing". en. In: *Proceedings of the AAAI Conference on Artificial Intelligence* 30.1 (Feb. 2016). Number: 1. ISSN: 2374-3468. DOI: 10.1609/aaai.v30i1.10139. URL: https://ojs.aaai.org/index.php/AAAI/article/view/10139 (visited on 10/30/2024).

[102]   Yong Yu et al. "A Review of Recurrent Neural Networks: LSTM Cells and Network Architectures". In: *Neural Computation* 31.7 (July 2019), pp. 1235–1270. ISSN: 0899-7667. DOI: 10.1162/neco_a_01199. URL: https://doi.org/10.1162/neco_a_01199 (visited on 10/30/2024).

[103]   Daya Guo et al. *GraphCodeBERT: Pre-training Code Representations with Data Flow*. arXiv:2009.08366. Sept. 2021. DOI: 10.48550/arXiv.2009.08366. URL: http://arxiv.org/abs/2009.08366 (visited on 10/30/2024).

[104]   Wasi Uddin Ahmad et al. *A Transformer-based Approach for Source Code Summarization*. arXiv:2005.00653. May 2020. DOI: 10.48550/arXiv.2005.00653. URL: http://arxiv.org/abs/2005.00653 (visited on 10/30/2024).

[105]   Yujia Li et al. *Gated Graph Sequence Neural Networks*. arXiv:1511.05493. Sept. 2017. DOI: 10.48550/arXiv.1511.05493. URL: http://arxiv.org/abs/1511.05493 (visited on 10/30/2024).

[106]   Yaqin Zhou et al. "Devign: Effective Vulnerability Identification by Learning Comprehensive Program Semantics via Graph Neural Networks". In: *Advances in Neural Information Processing Systems*. Vol. 32. Curran Associates, Inc., 2019. URL: https://proceedings.neurips.cc/paper_files/paper/2019/hash/49265d2447bc3bbfe9e76306ce40a31f-Abstract.html (visited on 10/30/2024).

[107]   Wasi Uddin Ahmad et al. *Unified Pre-training for Program Understanding and Generation*. arXiv:2103.06333. Apr. 2021. DOI: 10.48550/arXiv.2103.06333. URL: http://arxiv.org/abs/2103.06333 (visited on 10/30/2024).

[108]   Thomas N. Kipf and Max Welling. *Semi-Supervised Classification with Graph Convolutional Networks*. arXiv:1609.02907. Feb. 2017. DOI: 10.48550/arXiv.1609.02907. URL: http://arxiv.org/abs/1609.02907 (visited on 10/30/2024).

[109]   Zimin Chen et al. "SequenceR: Sequence-to-Sequence Learning for End-to-End Program Repair". In: *IEEE Transactions on Software Engineering* 47.9 (Sept. 2021). Conference Name: IEEE Transactions on Software Engineering, pp. 1943–1959. ISSN: 1939-3520. DOI: 10.1109/TSE.2019.2940179. URL: https://ieeexplore.ieee.org/abstract/document/8827954 (visited on 10/30/2024).

[110]   Qihao Zhu et al. *A Syntax-Guided Edit Decoder for Neural Program Repair*. arXiv:2106.08253. Mar. 2022. DOI: 10.48550/arXiv.2106.08253. URL: http://arxiv.org/abs/2106.08253 (visited on 10/30/2024).

[111]   Baptiste Roziere et al. "Code llama: Open foundation models for code". In: *arXiv preprint arXiv:2308.12950* (2023).

[112]   Hamel Husain et al. *CodeSearchNet Challenge: Evaluating the State of Semantic Code Search*. arXiv:1909.09436. June 2020. DOI: 10.48550/arXiv.1909.09436. URL: http://arxiv.org/abs/1909.09436 (visited on 10/30/2024).

[113]   Veselin Raychev, Pavol Bielik, and Martin Vechev. "Probabilistic model for code with decision trees". In: *Proceedings of the 2016 ACM SIGPLAN International Conference on Object-Oriented Programming, Systems, Languages, and Applications*. OOPSLA 2016. New York, NY, USA:

Association for Computing Machinery, Oct. 2016, pp. 731–747. ISBN: 978-1-4503-4444-9. DOI: 10.1145/2983990.2984041. URL: https://dl.acm.org/doi/10.1145/2983990.2984041 (visited on 10/30/2024).

[114] Bander Alsulami et al. "Source code authorship attribution using long short-term memory based networks: 22nd European Symposium on Research in Computer Security, ESORICS 2017". In: *Computer Security − ESORICS 2017 - 22nd European Symposium on Research in Computer Security, Proceedings*. Lecture Notes in Computer Science (including subseries Lecture Notes in Artificial Intelligence and Lecture Notes in Bioinformatics) (2017). Ed. by Einar Snekkenes, Simon N. Foley, and Dieter Gollmann. Publisher: Springer Verlag, pp. 65–82. ISSN: 9783319664019. DOI: 10.1007/978-3-319-66402-6_6. URL: http://www.scopus.com/inward/record.url?scp=85029521566&partnerID=8YFLogxK (visited on 10/30/2024).

[115] Hui-Hui Wei and Ming Li. "Supervised deep features for software functional clone detection by exploiting lexical and syntactical information in source code". In: *Proceedings of the 26th International Joint Conference on Artificial Intelligence*. IJCAI'17. Melbourne, Australia: AAAI Press, Aug. 2017, pp. 3034–3040. ISBN: 978-0-9992411-0-3. (Visited on 10/30/2024).

[116] CodeChef. *Codechef: Practical coding for everyone*. 2024. URL: https://codechef.com/.

[117] Xing Hu et al. "Summarizing source code with transferred API knowledge". In: *Proceedings of the 27th International Joint Conference on Artificial Intelligence*. IJCAI'18. Stockholm, Sweden: AAAI Press, July 2018, pp. 2269–2275. ISBN: 978-0-9992411-2-7. (Visited on 10/30/2024).

[118] Rjust. *RJUST/DEFECTS4J: A database of real faults and an experimental infrastructure to enable controlled experiments in Software Engineering Research*. 2024. URL: https://github.com/rjust/defects4j.

[119] Google. *Google Code Jam*. 2017. URL: https://code.google.com/.

[120] Beiqi Zhang et al. "Practices and challenges of using github copilot: An empirical study". In: *arXiv preprint arXiv:2303.08733* (2023).

[121] Mark Chen et al. *Evaluating Large Language Models Trained on Code*. arXiv:2107.03374. July 2021. DOI: 10.48550/arXiv.2107.03374. URL: http://arxiv.org/abs/2107.03374 (visited on 10/30/2024).

[122] GitHub. *GitHub copilot · your AI pair programmer*. 2024. URL: https://github.com/features/copilot.

[123] EleutherAI. *ELEUTHERAI/GPT-J-6B · hugging face*. 2024. URL: https://huggingface.co/EleutherAI/gpt-j-6b.

[124] Sid Black et al. "GPT-Neo: Large Scale Autoregressive Language Modeling with Mesh-Tensorflow". In: Version Number: 1.0. Zenodo, Mar. 2021. DOI: 10.5281/ZENODO.5297715. URL: https://zenodo.org/record/5297715 (visited on 10/30/2024).

[125] CodeParrot. *Codeparrot (CodeParrot)*. 2024. URL: https://huggingface.co/codeparrot.

[126] Erik Nijkamp et al. *CodeGen: An Open Large Language Model for Code with Multi-Turn Program Synthesis*. arXiv:2203.13474. Feb. 2023. DOI: 10.48550/arXiv.2203.13474. URL: http://arxiv.org/abs/2203.13474 (visited on 10/30/2024).

[127] Aohan Zeng et al. *GLM-130B: An Open Bilingual Pre-trained Model*. arXiv:2210.02414. Oct. 2023. DOI: 10.48550/arXiv.2210.02414. URL: http://arxiv.org/abs/2210.02414 (visited on 10/30/2024).

[128] Raymond Li et al. "Starcoder: may the source be with you!" In: *arXiv preprint arXiv:2305.06161* (2023).

[129] Anton Lozhkov et al. "Starcoder 2 and the stack v2: The next generation". In: *arXiv preprint arXiv:2402.19173* (2024).

[130] Mistral AI. *Codestral: Mistral's Code Model*. Accessed: 2025-06-05. 2024. URL: https://mistral.ai/news/codestral.

[131] Daya Guo et al. "DeepSeek-Coder: When the Large Language Model Meets Programming–The Rise of Code Intelligence". In: *arXiv preprint arXiv:2401.14196* (2024).

[132] Yuxiang Wei et al. "Magicoder: Empowering code generation with oss-instruct". In: *arXiv preprint arXiv:2312.02120* (2023).

[133] CodeGemma Team et al. "Codegemma: Open code models based on gemma". In: *arXiv preprint arXiv:2406.11409* (2024).

[134] Frank F Xu et al. "A systematic evaluation of large language models of code". In: *Proceedings of the 6th ACM SIGPLAN international symposium on machine programming*. 2022, pp. 1–10.

[135] Microsoft Research. *Phi-2: The surprising power of small language models*. Accessed: 2025-06-05. 2023. URL: https://www.microsoft.com/en-us/research/blog/phi-2-the-surprising-power-of-small-language-models/.

[136] Louis Kim and Rebecca Russell. *Draper VDISC dataset - vulnerability detection in source code*. Apr. 2024. URL: https://osf.io/d45bw/.

[137] Michele Tufano et al. "An Empirical Study on Learning Bug-Fixing Patches in the Wild via Neural Machine Translation". In: *ACM Trans. Softw. Eng. Methodol.* 28.4 (Sept. 2019), 19:1–19:29. ISSN: 1049-331X. DOI: 10.1145/3340544. URL: https://dl.acm.org/doi/10.1145/3340544 (visited on 10/30/2024).

[138] JiayangWu. *Jiayangwu/leetcode-python: LeetCode Solutions in python2. LeetCode in python*. 2023. URL: https://github.com/JiayangWu/LeetCode-Python.

[139] Jacob Austin et al. "Program Synthesis with Large Language Models". In: *International Conference on Learning Representations (ICLR)*. 2021. URL: https://paperswithcode.com/dataset/mbpp?utm_source=chatgpt.com.

[140] Nidhish Shah, Zulkuf Genc, and Dogu Araci. "StackEval: Benchmarking LLMs in Coding Assistance". In: *Advances in Neural Information Processing Systems* 37 (2024), pp. 36976–36994.

[141] Yujia Li et al. "Competition-level code generation with alphacode". In: *Science* 378.6624 (2022), pp. 1092–1097.

[142] Tao Yu et al. "Spider: A Large-Scale Human-Labeled Dataset for Complex and Cross-Domain Semantic Parsing and Text-to-SQL Task". In: *Proceedings of the 2018 Conference on Empirical Methods in Natural Language Processing (EMNLP)*. 2018. URL: https://arxiv.org/abs/1809.08887.

[143] Ruchir Puri et al. *CodeNet: A Large-Scale AI for Code Dataset for Learning a Diversity of Coding Tasks*. arXiv:2105.12655. Aug. 2021. DOI: 10.48550/arXiv.2105.12655. URL: http://arxiv.org/abs/2105.12655 (visited on 10/30/2024).

[144] Daniel Ramos et al. "Are Large Language Models Memorizing Bug Benchmarks?" In: *arXiv preprint arXiv:2411.13323* (2024).

[145] Bryce Goodman and Seth Flaxman. "European Union regulations on algorithmic decision-making and a "right to explanation"". In: *AI magazine* 38.3 (2017), pp. 50–57.

[146] Mohammad Mahdi Sayyadnejad et al. "Exploring the Black Box: Analyzing Explainable AI Challenges and Best Practices Through Stack Exchange Discussions". In: (2024).

[147] Yue Jia and Mark Harman. "An analysis and survey of the development of mutation testing". In: *IEEE transactions on software engineering* 37.5 (2010), pp. 649–678.

[148] Qianqian Zhu, Annibale Panichella, and Andy Zaidman. "A systematic literature review of how mutation testing supports quality assurance processes". In: *Software Testing, Verification and Reliability* 28.6 (2018), e1675.

[149] Nargiz Humbatova, Gunel Jahangirova, and Paolo Tonella. "Deepcrime: mutation testing of deep learning systems based on real faults". In: *Proceedings of the 30th ACM SIGSOFT international symposium on software testing and analysis*. 2021, pp. 67–78.

[150]  Annibale Panichella and Cynthia CS Liem. "What are we really testing in mutation testing for machine learning? a critical reflection". In: *2021 IEEE/ACM 43rd International Conference on Software Engineering: New Ideas and Emerging Results (ICSE-NIER)*. IEEE. 2021, pp. 66–70.

[151]  David Schuler and Andreas Zeller. "Covering and uncovering equivalent mutants". In: *Software Testing, Verification and Reliability* 23.5 (2013), pp. 353–374.